\documentclass[reqno]{amsart}

\usepackage{amssymb,amsxtra}
\usepackage{amsfonts}
\usepackage{amsmath}
\usepackage[english]{babel}
\usepackage{graphicx}
\usepackage{fancyhdr}
\usepackage{mathrsfs}
\usepackage{layout}

\newcommand{\nuu}{\frac{1-\nu}{1+\nu} }

\newcommand{\bdm}{\begin{displaymath}}
\newcommand{\edm}{\end{displaymath}}
\newcommand{\bdn}{\begin{eqnarray}}
\newcommand{\edn}{\end{eqnarray}}
\newcommand{\bay}{\begin{array}{c}}
\newcommand{\eay}{\end{array}}
\newcommand{\ben}{\begin{enumerate}}
\newcommand{\een}{\end{enumerate}}
\newcommand{\beq}{\begin{equation}}
\newcommand{\eeq}{\end{equation}}
\newcommand{\bra}[1]{\lf\langle #1 \ri|}
\newcommand{\ket}[1]{\lf|#1 \ri\rangle}
\newcommand{\braket}[2]{\lf. \lf\langle #1 \ri|  #2\ri\rangle}
\newcommand{\enne}{\mathbb{N}}
\newcommand{\erre}{\mathbb{R}}
\newcommand{\erd}{\mathbb{R}^d}
\newcommand{\erdd}{\mathbb{R}^{2d}}
\newcommand{\err}{\mathbb{R}^2}

\newcommand{\half}{\frac{1}{2}}

\newcommand{\de}{\delta}
\newcommand{\al}{\alpha}

\newcommand{\ga}{\gamma}
\newcommand{\la}{\lambda}
\newcommand{\fhi}{\varphi}

\newcommand{\piano}{\Pi}
\newcommand{\nw}{\newline}

\newcommand{\lf}{\left}
\newcommand{\ri}{\right}
\newcommand{\xp}{x^{\prime}}
\newcommand{\yp}{y^{\prime}}
\newcommand{\xv}{\vec{x}}
\newcommand{\yv}{\vec{y}}
\newcommand{\kv}{\vec{k}}
\newcommand{\xvp}{\vec{x}^{\prime}}
\newcommand{\yvp}{\vec{y}^{\prime}}
\newcommand{\xpq}{x^{\prime 2}}
\newcommand{\ypq}{y^{\prime 2}}

\providecommand{\absq}[1]{\lvert #1 \rvert^2}
\newcommand{\FO}{{ F}_0^{\omega}}
\newcommand{\Fa}{{F}_{\al}^{\omega}}
\newcommand{\Fia}{\Phi_{\al,\omega}^{\la}}
\newcommand{\DFO}{{\mathscr D}({F}^{\omega}_0)}
\newcommand{\DFa}{{\mathscr D}({ F}_{\al}^{\omega})}
\newcommand{\DFia}{\mathscr{D}(\Fia)}
\newcommand{\bml}{\begin{gather}}
\newcommand{\eml}{\end{gather}}
\newcommand{\ten}{\rightarrow}

\newcommand{\nt}{\noindent}
\newcommand{\n}{\noindent}

\newcommand{\spaz}{\vspace{.5cm} \nt}
\newcommand{\spazietto}{\vspace{.2cm} \nt}

\newcommand{\hlog}{\mathscr{H}^{\mathrm{log}}(\err)}
\newcommand{\gala}{\Gamma^{\la}}
\newcommand{\galao}{\Gamma^{\la}_{\omega}}

\newcommand{\gae}{\Gamma^{E}}
\newcommand{\Gt}{ {\mathcal G}^{\la}}
\newcommand{\Gta}{{\mathcal G}^{\la \ast}}
\newcommand{\Hr}{H^{\omega}_{\mathrm{osc}}}
\newcommand{\Hro}{H^{1}_{\mathrm{osc}}}
\newcommand{\I}{\mathcal P }

\newcommand{\Mx}{\langle x \rangle}
\newcommand{\Mp}{\langle p \rangle}
\newcommand{\Mk}{\langle k \rangle}

\newtheorem{teo}{Theorem}[section]
\newtheorem{dfn}[teo]{Definition}
\newtheorem{lem}[teo]{Lemma}

\newtheorem{prop}[teo]{Proposition}

\newenvironment{dem}{\vspace{.2cm}\nt {\bf Proof }\\}{\begin{flushright} $ \Box $ \end{flushright}}

\numberwithin{equation}{section}

\begin{document}

\title[Spectral analysis]{Spectral analysis of a two body problem with
zero range perturbation}

\author{M. Correggi}
\address{Michele Correggi}
\curraddr{Scuola Normale Superiore, P.zza dei Cavalieri 7, 56126 Pisa, Italy.}
\email{m.correggi@sns.it}

\author{G. Dell'Antonio}
\address{Gianfausto Dell'Antonio}
\curraddr{Dipartimento di Matematica, Universit\`a Di Roma ``La Sapienza'', P.za A.Moro 5, 00185 Rome, Italy.}
\email{dellantonio@mat.uniroma1.it}

\author{D. Finco} 
\address{Domenico Finco}
\curraddr{Dipartimento di Matematica, Universit\`a Di Roma ``La Sapienza'', P.za A.Moro 5, 00185 Rome, Italy.}
\email{finco@mat.uniroma1.it}

\begin{abstract} We consider a class of singular, zero-range 
perturbations of the
Hamiltonian of a quantum system composed by a test particle and a
harmonic oscillators in dimension one, two and three and we study
its spectrum. In facts we give a detailed characterization of point spectrum
and its asymptotic behavior with respect to the parameters entering the Hamiltonian. We also partially describe the positive spectrum and scattering
properties of the Hamiltonian.

\end{abstract}

\maketitle

\section{Introduction}

\nt
We consider in $ \erre^d $, $ d= 1,2,3 $, a system composed of a test particle  and a harmonic oscillator interacting through a zero-range force.
\nw
The hamiltonian is formally written as
\beq
	\label{ham}
	H_{\al}^{\omega} \equiv H_0^{\omega} 
	+ \mbox{``}\alpha \; \delta (\xv-\yv)\mbox{''}, \qquad H_0^{\omega} \equiv - \half \Delta_{\xv} 
- \half \Delta_{\yv} + \frac{\omega^2 y^2}{2} - \frac{\omega d}{2}  
\eeq
The precise meaning of the formal expression ``$  \alpha \; \delta(\xv-\yv) $'' will be given shortly.
\nw
Hamiltonians with formal zero-range forces have been introduced in Physics since the early '30 (see, e.g., \cite{fe,kp}), in particular for the study of scattering of low energy atoms and electrons from a target.
\nw
The operators often considered in literature are approximations of \eqref{ham},
which correspond, very roughly speaking, to interactions with ``very massive''
nuclei through a potential of ``very short range'' and the nuclei are supposed
so massive that they can be regarded as fixed scattering centers.
\nw
Here we consider a more general model in which the nuclei of the target are regarded as quantum particles harmonically  bound to their equilibrium positions. This kind of model is widely used in Physics to reconstruct the structure of the target (e.g., the distribution of the equilibrium positions) from scattering data.
\nw
In this paper we treat only the case of one harmonic oscillator, in which case $ \alpha $ is a real parameter. We will come back in a forthcoming paper to the case in which several oscillators are present.
\nw
From the point of view of mathematics,  zero-range interactions are interesting non-trivial models, for which it is possible to find simple explicit  solutions of the Schr\"{o}dinger equation and  compute physical relevant quantities. These models can be indexed by a small number of parameters which codify the ``strength'' 
and the position of the interactions.
\nw
The cases $d= 1,2,3$ will be treated in Sections 2, 3 and 4 respectively, and in each Section we shall recall some results about the rigorous definition of zero range interactions in the corresponding dimension, often referring to \cite{me} for proofs and further details.
\nw
Here we only remark that the definition of zero range interaction is much easier in dimension one
when it can be given in terms of boundary conditions at $x=y$.
\nw
On the contrary, in dimension two and three the definition requires a more sophisticated analysis in terms of self-adjoint extensions of the restriction of the operator $H_0^{\omega}$ to smooth functions which vanish in some neighborhood of the hyperplane $ \xv = \yv $. This extension may be obtained at a purely formal level, considering a suitable  interaction of range $ \epsilon $ and letting $ \epsilon \to 0 $, after having applied a suitable renormalization prescription. An equivalent rigorous definition is obtained through the theory of quadratic forms. This is the definition we shall use in our analysis, indeed we shall show that in dimension 1,2 and 3 the operators $ H_{\al}^{\omega} $ are determined by simple quadratic forms, closed and bounded below.
\nw
It is worth remarking that zero-range interactions, as we have defined them, do not exist when $ d  \geqslant 4 $ (the restriction mentioned above defines in this case an operator which is essentially self-adjoint) and that we consider only part of the self-adjoint extensions, i.e., those  commonly called ``$ \delta $''-type.
\nw
In the following Sections, we shall prove that the essential spectrum of $ H_{\al}^{\omega}$ is the half-line $ [0,+\infty) $, for all values of the parameters and $d = 1,2,3 $, and the wave operators $ \Omega_\pm  
(H_{\al}^{\omega}, H_0^{\omega})$ exist and are complete.
\nw
We shall also fully characterize the negative part of the spectrum  and give estimates of the number of eigenvalues.
\nw
We plan to come back to the scattering problem in a forthcoming paper and give a complete description of the multi-channel scattering associated to the pair $ H_{\al}^{\omega}, H _0^{\omega} $.

\subsection{Notation}

We introduce in this Section some notation and basic facts which will be used in the rest of the paper. 

\n
Vectors in $\erre^d$ will be denoted by $\vec{x}$, the modulus of $\vec{x}$ by 
$x$ and $\Mx$ stands for $(1 + x^2)^{1/2}$.

\nt
Unless stated otherwise $\| \cdot \|$ will denote both the norm of functions in $L^2(\erre^d)$ and the norm of
bounded endomorphism of $L^2(\erre^d)$.

\nt
Given any function $ f \in L^2(\erd) $, its Fourier transform, denoted by $ \hat{f} $, will be defined by
\beq
	\hat{f}(\kv) \equiv \frac{1}{(2\pi)^{\frac{d}{2}}} \int_{\erd} d\xv \: e^{-i\kv \cdot \xv} f(\xv).
\eeq
\nt
We shall denote the Sobolev space of order $ m $ by $ \mathcal{H}^{m}(\erd) $, i.e.,
\bdm
	 \mathcal{H}^{m}(\erd) \equiv \lf\{ f \in L^2(\erd) \: \Big| \: \Mk^m \hat{f}\in L^2(\erd) \ri\} \qquad \| f \|_{ \mathcal{H}^{m}} = \lf\| \Mk^m \hat{f} \ri\|
\edm
and the logarithmic Sobolev space by
\bdm
	\mathcal{H}^{\mathrm{log}}(\erd) \equiv \lf\{ f \in L^2(\erd) \: \Big| \: \log(1+\Mk) \hat{f} \in L^2(\erd) \ri\}
	\qquad \| f \|_{ \mathcal{H}^{\mathrm{log}}} = \lf\| \log(1+\Mk )\hat{f} \ri\|.
\edm
\n
We introduce the Hamiltonian of the harmonic oscillator by $\Hr = \frac{1}{2} ( p^2 + \omega^2 x^2 )$, which will be used as a reference Hamiltonian in some technical estimates, and denote by $ \Psi_n^{(\omega)}(\xv) $, $ n \in \enne $, its normalized eigenvectors. The integral kernel of the semigroup (Mehler kernel) is given by
\beq
\label{Mehler}
	e^{-\Hr t}(\xv;\xvp) \equiv \frac{e^{-\frac{\omega d t}{2}} }{\pi^{\frac{d}{2}} \lf( 1 - e^{-2 \omega t} \ri)^{\frac{d}{2}}} \exp \lf\{ - \frac{\omega ( x^2 + {\xp}^2 )}{2 \tanh \omega t} + \frac{\omega \xv \cdot \xvp}{\sinh \omega t} \ri\}.
\eeq  

\nt
Let us recall for the reader's convenience some facts about compact operators. Let 
${\mathscr H}$ and ${\mathscr H}'$ be two Hilbert spaces, we shall denote the space
of compact operators from ${\mathscr H}$ to ${\mathscr H}'$ by 
${\mathscr B}_{0} ( {\mathscr H},{\mathscr H}') $; if $A \in {\mathscr B}_{0} ( {\mathscr H},{\mathscr H}') $,
we denote by $\mu_n(A)$ its singular values with decreasing ordering $\mu_0 (A) \geqslant \mu_1 (A) \geqslant 
\ldots \geqslant 0$. For $1\leqslant p \leqslant \infty$, we shall denote the Schatten ideals\footnote{The norm $ \| A \|_1 $ will also be denoted by $ \mathrm{Tr}(|A|) $, the usual trace class norm.} by 
\begin{equation*}
{\mathscr B}_{p} ( {\mathscr H},{\mathscr H}') = \lf\{ A\in {\mathscr B}_{0} ( {\mathscr H},{\mathscr H}') \, \bigg| \,
\sum_{n=1}^{\infty} (\mu_{n}(A) )^p < +\infty \ri\} \qquad 
\| A\|_{ p } = \lf( \sum_{n=0}^{\infty} (\mu_{n}(A) )^p  
\ri)^{\frac{1}{p} },
\end{equation*}
and, for $p=\infty $, we simply have ${\mathscr B}_{\infty} ( {\mathscr H},{\mathscr H}') = 
{\mathscr B} ( {\mathscr H},{\mathscr H}')$ and $\|A\|_{\infty} = \|A\|$. Let us recall also 
that, for  $A \in {\mathscr B}_{p} ( {\mathscr H},{\mathscr H}') $, we have 
$\mu_{n}(A) =  \mu_{n}(A^{\ast}) $, where $\ast$ denotes the adjoint, and $ A^* \in {\mathscr B}_{p} ( {\mathscr H}',{\mathscr H}) $ (see, e.g., \cite{k}). We shall denote the spectrum of
an operator $A$ by $\sigma (A)$, the pure point spectrum by $\sigma_{\mathrm{pp}}(A)$ and the essential spectrum
by $\sigma_{\mathrm{ess}}(A)$.

\n
The resolvent of the operator  
\bdm
	H_0^{\omega} \equiv - \frac{1}{2} \Delta_{\xv} - \frac{1}{2} \Delta_{\yv} + \frac{\omega^2 y^2}{2} - \frac{\omega d}{2} ,
\edm
is given by the following integral kernel\footnote{In the following we shall often omit the suffix $ \omega $ and set $ G^{\la} \equiv G^{\la}_1 $. Similarly we denote by $ H_{\al} $ and $ H_0 $ the operators $ H_{\al}^{1} $ and $ H_0^1 $ respectively.},
\begin{multline}
	\label{gre}
	G_{\omega}^{\la}(\xv,\yv;\xvp,\yvp) \equiv \lf( H_0^{\omega} + \la \ri)^{-1}(\xv,\yv;\xvp,\yvp) =	\\
	= \frac{\omega^{d-1}}{2^{^\frac{d}{2}} \pi^{d}} \int_0^1 d\nu \: \frac{\nu^{ \frac{\la}{\omega} - 1 }}{\lf( 1- \nu^2 \ri)^{\frac{d}{2}} \lf( \ln \frac{1}{\nu} \ri)^{\frac{d}{2}}} \exp \lf\{ -\frac{\omega}{2} \nuu  (y^2 + \ypq) -   \frac{\omega}{ 2\ln {\frac{1}{\nu}  }  } (\xv-\xvp)^2 - \frac{ \omega \nu}{ 1 - \nu^2}  ( \yv - \yvp )^2 \ri\},
\end{multline}
where $ \la > 0 $ and $ \xv,\yv,\xvp,\yvp \in \erd $. The above expression has been obtained in \cite{me}. Note that in the one-dimensional case the kernel \eqref{gre} can be expressed as well as
\beq
	\label{gre2}
	G_{\omega}^{\la}(x,y;\xp,\yp) = \sum_{n=0}^{+\infty} \frac{\Psi^{(\omega)}_n(y) \Psi^{(\omega)}_n(\yp)}{\sqrt{2(\omega n+\la)}} \exp \lf\{ -\sqrt{2(\omega n+\la)} \lf| x-\xp \ri| \ri\}. 
\eeq 
The symbol $ \Pi $ will stand for the hyperplane
\beq
	\Pi \equiv \lf\{ (\xv,\yv) \in \erdd \: \big| \: \xv = \yv \ri\},
\eeq
and $ \Gt_{\omega} f (\xv,\yv) $ for the potential associated with $ f \in L^2(\erd)$, i.e., 
\beq
\Gt_{\omega} f (\xv,\yv) \equiv \int_{\erd} d\xvp \; G_{\omega}^{\la} (\xv,\yv;\xvp,\xvp) f(\xvp).
\eeq

\n
If we denote by $\I : {\mathcal H}^{m}(\erre^{2d}) \ten L^2(\erre^d) $, $m>d/2$, the restriction to the plane 
$\piano$, we trivially have  $\Gt = G_{\omega}^{\la} \I^{\ast}$.

\nt
Any positive constant will be denoted by $ c $, whose value may change from line to line.

\section{The One-dimensional Case}

\subsection{Preliminary Results}	

The easiest way to give the expression \eqref{ham} a rigorous meaning is to consider the (formal) quadratic form associated with such an operator: For any $ \al \in \erre $, at least formally, we have
\bdm
	\bra{u} H_{\al}^{\omega} \ket{u}  = \bra{u} H_0^{\omega} \ket{u} + \al \int_{\erre} dx \lf| u(x,x) \ri|^2.
\edm 
This formal expression identifies a closed quadratic form bounded below (see \cite{me} for the proof):

\spazietto
\begin{dfn}[Quadratic Form $ \Fa $]
	\mbox{}	\\
	The quadratic form $\lf( \Fa, \DFa \ri)$ is defined as follows,
	\beq
	 	\label{forma1d}
		\Fa[u] \equiv \int_{\err} dx dy \lf\{ \half \lf| \frac{\partial u}{\partial x} \ri|^2 +  
 		\half \lf| \frac{\partial u}{\partial y} \ri|^2 + \frac{\omega^2  y^2}{2}\absq{u }  -\frac{\omega}{2}  \absq{u }\ri\}
		+ \al  \int_{\erre} dx\, \absq{ u(x,x) } \equiv \FO[u] + \al { F}_{\mathrm{int}} [u], 
	\eeq
	\begin{equation}
		\label{lampacione}
		\DFa= \lf\{ u \in L^2( \erre^{2} )\: | \: \Fa 
		[u] < +\infty \ri\}.
	\end{equation}
\end{dfn} 
 
\nt
The main properties of $\Fa$ are summarized in the following
\spazietto

\begin{teo}[Closure of the Form $ \Fa $]
	\mbox{}	\\
	The quadratic form $\lf( \Fa, \DFa \ri)$ is closed and bounded below on 
	\bdm
		\DFa = \DFO= \lf\{ u\in L^2( \erre^{2} )\: | \:  u \in H^1(\erre^{2}),\; y u \in L^2( \erre^{2} ) \ri\}.
	\edm
	\label{teopaguro}
\end{teo}

\nt
We denote by $(H_{\al}^{\omega}, {\mathscr D}(H_{\al}^{\omega}))$ the bounded from below self adjoint
operator in $L^{2}(\erre^{2})$ defined by
\eqref{forma1d} and \eqref{lampacione}. 
Concerning the resolvent of $H_{\al}^{\omega}$ we have the following result:
\spazietto

\begin{teo}[Operator $ H_{\al}^{\omega} $]
	\label{teocozza}
	\mbox{}	\\
	The domain and the action of $H_{\al}^{\omega}$ are the following,
	\begin{equation}
		{\mathscr D}(H_{\al}^{\omega}) = \lf\{ u \in L^2(\erre^{2} ) \: \big| \: u=\fhi^{\la}+ \Gt_{\omega} q, \, \fhi^{\la}\in {\mathscr D}(H_{0}^{\omega}),\,
		q + \al \:\I  u =0\ri\},
		\label{res1d2p}
	\end{equation}
	\begin{equation}
		(H_{\al}^{\omega} + \la)u = (H_0^{\omega} +\la )\fhi^{\la}.
		\label{res1d2pp}
	\end{equation}
	Moreover there exists $\la_0>0$  such that, for $\la > \la_0 $ and for any $f \in L^2(\erre^{2} )$, one has
	\begin{equation}
		\label{res1d}
		 \lf( H_{\al}^{\omega}+\la \ri)^{-1} f  = \lf( H_0^{\omega} + \la \ri)^{-1} f  + \Gt_{\omega} q_f ,
	\end{equation}
	\nt
	where the charge $q_{f} $ is a solution of the following equation,
	\begin{equation}
		\label{res1dcharge}
		q_f +\al \lf\{  K^{\la}_{\omega} q_f   + \I \, G^{\la}_{\omega} f  \ri\} = 0 
	\end{equation}
	and $K^{\la}_{\omega} \equiv \I\, G^{\la}_{\omega} \, \I^{\ast}$ has integral kernel $K^{\la}_{\omega}(x; \xp) \equiv G^{\la}_{\omega}(x,x; \xp,\xp)$.
\end{teo}

\nt
It is straightforward to see that $H_{\al}^{\omega}$ is an extension of $\widetilde{H_0}$ defined by
\begin{equation*}
	{\mathscr D}(\widetilde{H_0} )= \lf\{ u \in C_0^{\infty}(\erre^{2} \setminus  \piano ) \ri\},
\end{equation*}
\begin{equation*}
	\widetilde{H_0}u = \lf[ - \half \Delta_{x} -\half \Delta_{y} + \frac{\omega^2 y^{2}}{2}   - \frac{\omega}{2} \ri]u.
\end{equation*}
Then, by definition, $H_{\al}^{\omega}$ is a perturbation of $H_0^{\omega}$ supported by the null set $\piano$, 
i.e., a rigorous counterpart of \eqref{ham}. It follows from a general argument (see Lemma C.2 
in \cite{al}) that $H_{\al}^{\omega}$ is a local operator, i.e., if $u=0$ in an open set $\Omega$, then $H_{\al}^{\omega}u=0$ 
in  $\Omega$.

\nt
The effect of the interaction is equivalent to the boundary condition $ q + \al \: \I u =0 $ satisfied
by $u\in {\mathscr D}(H_{\al}^{\omega})$ (see \eqref{res1d2p}) which uniquely fix the self adjoint 
extension of $\widetilde{H_0}$. Such a boundary condition is manifestly local, i.e., the value of $q$ at a given point $ x \in \erre $ is proportional to the value of $ u $ at the point $ (x,x) $.
In this sense the constructed hamiltonian $H_{\al}^{\omega}$ defines a local zero-range interaction.

\nt
Note that the quadratic form \eqref{forma1d} is not the most general zero-range perturbation of $F_0^{\omega}$:
It is clear, for instance, that, if we take a real function $\al(x)$ such that $\al \in L^{\infty}$, then
\beq
		{F}^{\prime}_{\al} [u] \equiv \int_{\err} dx dy \lf\{ \half \lf| \frac{\partial u}{\partial x} \ri|^2 +  
 		\half \lf| \frac{\partial u}{\partial y} \ri|^2 + \frac{\omega^2  y^2}{2}\absq{u }  -\frac{\omega}{2}  \absq{u }\ri\}
		+  \int_{\erre} dx\, \al(x) \absq{ u(x,x) } 
\eeq
define another zero-range perturbation of $F_0^{\omega}$.

\nt
It is straightforward to see that the 
boundary condition corresponding to $	{F}^{\prime}_{\al}$ is $ q(x) + \al(x) \: (\I u)(x) =0 $. The perturbation we use is distinguished by its 
invariance under translations along the coincidence manifold $ \piano $. In facts, the quadratic form \eqref{forma1d} gives the simplest ``$\de$"-like zero-range perturbation of $F_0^{\omega}$ which correspond to a local boundary condition. Similar remarks hold for the two and three dimensional
case too.

\subsection{Spectral Analysis}

\nt
In this section we shall study the spectrum of \eqref{ham}.
\newline
We analyze first the properties of the operator $ K^{\la}_{\omega} $, for fixed $ \omega = 1$:
\spazietto

\begin{prop}[Spectral Analysis of $ K^{\la}_{\omega} $]
	\mbox{}	\\
	The operator $K^{\la}_1: L^2(\erre^3) \ten L^2(\erre^3)$ is  compact, positive definite and self-adjoint. 
	\newline
	Let $\mu_n(\la)$, $n \in \enne$, be its eigenvalues arranged in a decreasing order: $\mu_n(\la)$  is a decreasing function of $\la$, for any $ n \in \enne $, and $ \lim_{\la \ten 0} \mu_0(\la) = +\infty $, whereas $ \lim_{\la \ten 0} \mu_n(\la) < +\infty $, for $ n > 0 $. 
	\newline
	Furthermore the following estimate holds true 
	\beq
	\label{carlone}
		\lf| \mu_n(\la) - \frac{1}{\sqrt{2(n+\la)}} \ri| \leqslant \frac{3}{5} \sqrt{\frac{2}{\pi}}
	\eeq 
	for any $ n \in \enne $.
	\label{opek}
\end{prop}

\begin{dem}
	In order to simplify the notation, we shall denote by $ K^{\la} $ the operator $ K^{\la}_1 $, i.e., $ K^{\la}_{\omega} $ for fixed $ \omega = 1 $.
	\newline
	Using the boundedness criterium for integral operators, see \cite{hs}, it is straightforward to see that
	\begin{equation}
		\| K^{\la} \| \leqslant c \int_0^1 d\nu \frac{\nu^{\la -1}}{\sqrt{1 -\nu^2 + \ln  \frac{1}{\nu}} }.
		\label{normak}
	\end{equation}
	Estimate \eqref{normak} implies that $\| K^{\la} \| \leqslant c \la^{-\frac{1}{2}} $ for $\la \ten 0$ and
	for $\la \ten +\infty$. The operator $K^{\la}$ is manifestly self-adjoint. 

	\nt
	We introduce the following decomposition
	\begin{equation}
		K^{\la}(x; \xp) = \int_0^1 d\nu \, m^{\la}(\nu) \: k_{\nu}(x; \xp),
	\label{pastiera}
	\end{equation}
	\begin{equation}
	\label{pastiera2}
		m^{\la}(\nu) \equiv
		\frac{\nu^{  \la  -1} }{ \sqrt{2}\pi
		\sqrt{1- \nu^2} \sqrt{\ln{  \frac{1}{\nu}  }} }, \hspace{0,5cm}
		k_{\nu}(x; \xp) \equiv
		\exp \lf\{ -\half \nuu  (x^2 + \xpq) -   \frac{(x-\xp)^2}{ 2\ln {\frac{1}{\nu}  }  }  - \frac{ \nu( x - \xp )^2}{ 1 - \nu^2}   \ri\}.
	\end{equation}
	Since $k_{\nu}$ is a positive operator valued function and $m^{\la}(\nu)$ is a positive function for $\nu \in (0,1)$, the operator $K^{\la}$ is positive and has empty kernel.

	\nt 
	Let us prove now that $K^{\la}$ is a compact operator: Using Lemma at pg. 65 in \cite{rs3}, 
	it is straightforward to prove that 
	\bdm
		\int_0^{1-\de} d\nu \; m^{\la}(\nu) \: k_{\nu}(x;\xp) 
	\edm
	is the kernel of a trace class operator, for any $ \delta < 1 $. 
	By Halmos criterium, it converges to $K^{\la}$ in the uniform topology, therefore $K^{\la}$ is  
	compact and positive.

	\nt
	By \eqref{pastiera} we have that $K^{\la_1} > K^{\la_2}$ for $\la_1 < \la_2$, 
	therefore the eigenvalues are decreasing functions of $\la$ by
	Min-Max Theorem (see, e.g., Theorem XIII.1 in \cite{rs4}).

	\nt
	In order to study the behavior of the eigenvalues of $K^{\la}$ for $\la \ten 0$, it suffices to notice that,
	if $f$ is orthogonal to $\exp \lf\{ -\frac{y^2}{2} \ri\}$, then $\lim_{\la \ten 0} \bra{f} K^{\la} \ket{f} < +\infty$; the statement
	follows by Min-Max Theorem and estimate \eqref{normak}.

	\nt
	We prove now the eigenvalue upper bound: Note first that (see, e.g., \cite{y})
	\bdm
		\lf( \Hro + \la - 1/2 \ri)^{-\frac{1}{2}} = \frac{1}{\sqrt{\pi}} \int_{0}^{\infty} dt \: \frac{e^{-\lf( \la - \frac{1}{2} \ri)t} e^{-\Hro t}}{\sqrt{t}},
	\edm
	which together with \eqref{Mehler} yields
	\bdm
		\lf( \Hro + \la - 1/2 \ri)^{-\frac{1}{2}} (x;\xp) = \frac{1}{\pi} \int_0^1 d\nu \: \frac{\nu^{  \la  -1} }{ \sqrt{2}\pi
		\sqrt{1- \nu^2} \sqrt{\ln{  \frac{1}{\nu}  }}} \exp \lf\{ -\half \nuu  (x^2 + \xpq) - \frac{ \nu}{ 1 - \nu^2}  ( x - \xp )^2 \ri\}.
	\edm 
	In order to apply Schur test (see \cite{hs}) to the operator $ \lf( \Hro + \la - 1/2 \ri)^{-\frac{1}{2}} - \sqrt{2} K^{\la} $, we estimate
	\begin{multline*}
		\int_{\erre} d\xp \: \lf\{ \lf( \Hro + \la - 1/2 \ri)^{-\frac{1}{2}} (x;\xp) - \sqrt{2} K^{\la}(x;\xp) \ri\} = \frac{1}{\sqrt{\pi}} \int_{0}^1 d\nu \: \frac{\nu^{\la-1}}{\sqrt{1+\nu^2} \sqrt{\ln \frac{1}{\nu}}} \exp \lf[ -\frac{(1-\nu^2)x^2}{2(1+\nu^2)} \ri] \cdot	\\
		\lf\{ 1 - \sqrt{\frac{(1+\nu^2)\ln \frac{1}{\nu}}{1-\nu^2 + (1+\nu^2)\ln \frac{1}{\nu}}} \exp \lf[ - \frac{(1-\nu)^3 x^2}{2(1+\nu^2)[1-\nu^2+(1+\nu^2)\ln \frac{1}{\nu}]} \ri] \ri\} \leqslant	\\
		\frac{1}{2 \sqrt{\pi}} \int_{0}^1 d\nu \: \frac{\nu^{\la-1}(1-\nu)^3}{\sqrt{\ln \frac{1}{\nu}} \: (1+\nu^2)^{\frac{3}{2}} \lf[1-\nu^2+(1+\nu^2)\ln \frac{1}{\nu}\ri]},
	\end{multline*}
	where we have used the inequality 
	\bdm
		\exp \lf\{-a_1 x^2 \ri\} - b \exp \lf\{-a_2 x^2 \ri\} \leqslant \frac{b(a_2-a_1)}{a_1} \exp \lf\{ -a_2 x^2 \ri\} \leqslant \frac{a_2-a_1}{a_1},
	\edm
	which holds true for any $ 0 < a_1 < a_2 $,  $ 0 < b < 1 $  and $ b a_2 > a_1 $. The last integral can be easily estimated by
	\bdm
		\int_{0}^1 d\nu \: \frac{\nu^{\la-1}(1-\nu)^3}{\sqrt{\ln \frac{1}{\nu}} \: (1+\nu^2)^{\frac{3}{2}} \lf[1-\nu^2+(1+\nu^2)\ln \frac{1}{\nu}\ri]} \leqslant \int_0^{\infty} dt \: \frac{\lf( 1 - e^{-t} \ri)^3}{t^{\frac{3}{2}}} \leqslant \int_0^1 dt \: t^{\frac{3}{2}} + \int_1^{\infty} dt \: t^{-\frac{3}{2}} \leqslant \frac{12}{5},
	\edm
	so that, using the kernel symmetry, one has
	\beq
	\label{carlone2}
		\lf\|  \lf( \Hro + \la - 1/2 \ri)^{-\frac{1}{2}} - \sqrt{2} K^{\la} \ri\| \leqslant \frac{6}{5\sqrt{\pi}}.
	\eeq
	The result is thus a simple consequence of Min-Max Theorem.
\end{dem}

\nt
We are now able to study the point spectrum of $H_{\al}^{\omega}$:
\spazietto

\begin{teo}[Negative Spectrum of $H_{\al}^{\omega}$]
	\label{spectral1d}
	\mbox{}	\\
	For $\al \geqslant 0$, $H_{\al}^{\omega}$ has no negative eigenvalues, while, for $\al<0$, there is a finite number $N_{\omega}(\al)$ of negative eigenvalues $-E_0(\al,\omega) \leqslant -E_1(\al,\omega) \leqslant \ldots\leqslant 0 $ satisfying the scaling property
	\beq
	\label{scaling}
		E_n(\al,\omega) = \omega E_n(\al/\sqrt{\omega},1).
	\eeq
	The corresponding eigenvectors are given by $ u_n = {\mathcal G}^{E_n}_{\omega}  q_n $, where $ q_n $ is a solution\footnote{Such a solution is actually unique once the $L^2-$norm of $ q_n $ is fixed, as it is by the $L^2-$normalization of $ u_n $.} of the homogeneous equation $ q_n + \al K^{E_n}_{\omega} q_n = 0 $.
	\newline
	Furthermore there exists $ \alpha_0 > 0 $ such that, for $ - \al_0 < \al < 0 $, $N_{\omega}(\al)=1$, whereas, for $| \al | \geqslant \alpha_0 $, $ N_{\omega}(\al) > 1 $. For fixed $ \omega > 0 $, the ground state energy $E_0(\al,\omega)$
	satisfies the asymptotics
	\beq
	\label{asyal0}
		E_0(\al,\omega) \sim \frac{\al^2}{2}
	\eeq	
	as $ \al \to 0 $. 	
\end{teo}

\begin{dem}
	First we derive an integral equation equivalent to the eigenvalue problem. 
	Let $u$ be a solution of $H_{\al}^{\omega} u = -E u$, $E>0$; using \eqref{res1d2pp}, this proves to be equivalent to
	\begin{equation*}
		\lf( H_0^{\omega} + E \ri) \phi^{\la} = (\la -E) \: \Gt_{\omega} q.
		\label{mollica}
	\end{equation*}

	\nt
	The first resolvent identity yields from \eqref{mollica} 
	\begin{equation*}
		\phi^{\la} = {\mathcal G}^{E}_{\omega} q - \Gt_{\omega} q,
	\end{equation*}
	which implies 
	\begin{equation*}
		u = {\mathcal G}^{E}_{\omega}  q.
	\end{equation*}
	On the other hand by using the boundary conditions in \eqref{res1d2p}, we arrive at the following homogeneous equation for $q$ and $E$:
	\begin{equation}
		q + \al K^{E}_{\omega} q = 0.
		\label{statleg}
	\end{equation}
	By scaling the above equation is equivalent to the following one
	\beq
	\label{statlegsc}
		\tilde{q} + \frac{\al}{\sqrt{\omega}} K^{E/\omega}_{1} \tilde{q} = 0
	\eeq
	i.e., $ q$ solves \eqref{statleg} if and only if
	\bdm
		\tilde{q}(x) \equiv \omega^{-\frac{1}{4}} \: q(x/\sqrt{\omega})
	\edm
	solves \eqref{statlegsc}, which implies \eqref{scaling}.	
	\newline
	The other properties of the negative eigenvalues follows then from Proposition \ref{opek}: If $\al \geqslant 0$, \eqref{statlegsc} has no solution, since $K^{E/\omega}_1$ is a positive operator; if $\al<0$, by projecting \eqref{statlegsc} onto the eigenvectors of $ K^{E/\omega}_1$, one obtains the algebraic equation 
	\begin{equation}
		1+ \frac{\al \mu_n(E/\omega)}{\sqrt{\omega}} =0,
		\label{statleg2}
	\end{equation}
	and the eigenvalue equation is equivalent to find some $n \in \enne$ and $E > 0$ satisfying \eqref{statleg2}.
	\newline
	The monotonicity of $ \mu_n $ together with their asymptotics as $ E \to 0 $ (see Proposition \ref{opek}) imply that, for any $\al <0$, \eqref{statleg2} has only a finite number of solutions $E_n$. More precisely, for $| \al | < \al_0 $, it has only one solution $E_0$, since $\lim_{E \ten 0} \mu_0(E/\omega) = +\infty$ and $\lim_{E \ten 0} \mu_n(E/\omega) < +\infty$, for $ n > 0 $. 
	\newline
	For fixed $ \omega $, the estimate \eqref{carlone2} yields
	\beq
	\label{eigenasym}
		\mu_0(E/\omega,1) = \sqrt{\frac{\omega}{2E}} + \mathcal{O} \lf( \sqrt{E} \ri)
	\eeq
	as $ E \to 0 $ and Eq. \eqref{asyal0} easily follows from \eqref{statleg2}. 
\end{dem}

\nt
Note that the result contained in Theorem above yields also the expected asymptotic behavior as $ \omega \to 0 $: The limiting system is given by two particles freely moving on the line with a mutual zero-range interaction. The spectrum of such an operator is absolutely continuous for any sign of $ \al $, because of the translation invariance associated with the motion of the center of mass and it is $[0, \infty)$ for $\al> 0$, $[-\al^2 /2 ,\infty)$ for $\al<0$
. If $ \al < 0 $, the scaling property \eqref{scaling} implies that the eigenvalues accumulate at the bottom of the continuous spectrum as $ \omega \to 0 $ and the corresponding bound states eventually disappear.

\nt
More interesting is the opposite asymptotics, that is the limit $ \omega \to \infty $: In this case the strength of the harmonic oscillator becomes so large that, roughly speaking, one of the two particles remains fixed at the origin. More precisely we expect that the reduced dynamics of the other particle is generated by an hamiltonian formally given by 
\bdm
	h_{\al} = - \frac{1}{2} \Delta_x + \mbox{``} \alpha \delta(x) \mbox{''}.
\edm
We define $h_{\al}$ as the self adjoint operator
corresponding to the closed and bounded below quadratic form $f_{\al}$ given by
\begin{equation*}
f_{\al}[u]= \frac{1}{2} \int_{\erre} dx \, \lf| \frac{du}{dx} \ri|^2 + \al |u(0)|.
\end{equation*}
It is straightforward to compute the domain and the action of $h_{\al}$ (see, e.g., \cite{al}):
\beq
	h_{\al} = - \frac{1}{2} \frac{d^2}{dx^2},
\eeq
\beq
	\mathscr{D}(h_{\al}) = \left\{ u \in H^1(\erre) \cap H^2(\erre \setminus \{0\}) \: | \: u^{\prime}(0^+) - u^{\prime}(0^-) = 2 \al u(0) \right\},
\eeq 
and, for $ \al < 0 $, its spectrum contains only one negative eigenvalue $ - \alpha^2/2 $ with (normalized) eigenvector 
\bdm
	\xi_{\al}(x) \equiv \sqrt{|\al|} \: e^{-|\al| |x|}.
\edm
For fixed $ \al < 0 $ and $ \omega $ sufficiently large (larger than $ \al^2/2 $), the operator $ H_{\al}^{\omega} $ has only one negative eigenvalue $ E_0(\al,\omega) $ with normalized eigenvector $ u_{\al,\omega}(x,y) $. We denote by $ \rho_{\al,\omega} $ the reduced density matrix associated with the ground state $ u_{\al,\omega}(x,y) $, i.e., the trace class operator $ \rho_{\al,\omega}  : L^2(\erre) \to L^2(\erre) $ with integral kernel
\beq
	\rho_{\al,\omega}(x;\xp) \equiv \int_{\erre} dy \: u^*_{\al,\omega}(x,y) u_{\al,\omega}(\xp,y).
\eeq 
\spazietto 

\begin{prop}[Ground State Asymptotics as $ \omega \to \infty $]
	\mbox{}	\\
	For any fixed $ \al  < 0 $ and for $ \omega \to \infty $, 
	\beq
	\label{asyomegaeig}
		E_0(\al,\omega) = \frac{\al^2}{2} + \mathcal{O} \lf( \omega^{-1} \ri),
	\eeq
	and the reduced density matrix $ \rho_{\al,\omega} $ converges to the one-dimensional projector onto $ \xi_{\al} $, i.e.,
	\beq
	\label{asyomegamat}
		\rho_{\al,\omega} \underset{\omega \to \infty}{\longrightarrow} \lf| \xi_{\al} \rangle \langle \xi_{\al} \ri|.
	\eeq
	in the norm topology of $ \mathscr{B}_1(L^2(\erre)) $.
\end{prop}

\begin{dem}
	We first notice that the bound $\| K^{\la} \| \leqslant c \la^{-\frac{1}{2}} $ (see \eqref{normak}) together with the eigenvalue equation \eqref{statleg2} imply the bound $ E_0(\al,\omega) \leqslant c \al^2 $, so that the ground state energy of $ H_{\al}^{\omega} $ is bounded (from above) uniformly in $ \omega $ and $ E_0/\omega \to 0 $, as $ \omega \to \infty $, for fixed $ \al $. Therefore the first part of the statement can be proved exactly as the asymptotics \eqref{asyal0} (see, e.g., \eqref{eigenasym}). 
	\newline
	We now consider the ground state wave function $ u_{\al,\omega} $, which can be expressed as $ u_{\al,\omega} = {\mathcal G}^{E_0}_{\omega}  q_0 $ (see Proposition \ref{spectral1d}), where $ q_0 $ is a solution of the homogeneous equation $ q_0 + \al K^{E_0}_{\omega} q_0 = 0 $. Note that the $L^2-$norm of $ q_0 $ is actually fixed by the normalization of $ u_{\al,\omega} $. Let us decompose $ q_0 $ as $ q_0 = Q_0 \Psi_0^{(\omega)} + \xi $, with $ \braket{\Psi_0^{(\omega)}}{\xi} = 0 $. We are going to prove that
	\beq
	\label{asygstate}
		\lf\| u_{\al,\omega} - w_{\al,\omega} \ri\|^2 \leqslant \frac{c \| q_0 \|^2}{\omega^{\frac{3}{2}}}, \hspace{1,5cm} w_{\al,\omega}(x,y) \equiv \frac{Q_0}{\sqrt{2E_0}} \Psi_0^{(\omega)}(y) \exp \lf\{-\sqrt{2E_0}|x| \ri\}.
	\eeq
	The proof is done in two steps: We first show that
	\bdm
		\lf\| u_{\al,\omega} - v_{\al,\omega} \ri\|^2 \leqslant \frac{c \| q_0 \|^2}{\omega^{\frac{3}{2}}}, \hspace{1,5cm} v_{\al,\omega}(x,y) \equiv \frac{1}{\sqrt{2E_0}} \Psi_0^{(\omega)}(y) \int_{\erre} d\xp \: e^{-\sqrt{2E_0}|x-\xp|} \Psi_0^{(\omega)}(\xp)\:  q_0(\xp).
	\edm
	Indeed, by using the representation \eqref{gre2}, we can easily estimate
	\bdm
		\lf\| u_{\al,\omega} - v_{\al,\omega} \ri\|^2 \leqslant \sum_{n=1}^{\infty} \frac{1}{\sqrt{2} (\omega n + E_0)^{\frac{3}{2}}} \lf| \braket{\lf|\Psi_n^{(\omega)}\ri|}{|q_0|} \ri|^2 \leqslant \frac{c \| q_0 \|^2}{\omega^{\frac{3}{2}}} \sum_{n=1}^{\infty} \frac{1}{n^{\frac{3}{2}}} \leqslant \frac{c \| q_0 \|^2}{\omega^{\frac{3}{2}}}.
	\edm 
	On the other hand, setting $ \tilde{q}_0(x) \equiv \omega^{-\frac{1}{4}} q_0(x/\sqrt{\omega}) $, one has
	\beq
	\label{wnorm}
		\lf\| w_{\al,\omega} \ri\|^2 = \frac{|Q_0|^2}{(2E_0)^{\frac{3}{2}}}, 
	\eeq
	\bdm
		\lf\| v_{\al,\omega} \ri\|^2 = \frac{|Q_0|^2}{(2E_0)^{\frac{3}{2}}} + \frac{1}{2E_0 \omega^{\frac{3}{2}}} \int_{\erre} dx \int_{\erre} d\xp \: |x-\xp| \Psi_0^{(1)}(x) \Psi_0^{(1)}(\xp) \tilde{q}_0^*(x) \tilde{q}_0(\xp) \leqslant  \frac{|Q_0|^2}{(2E_0)^{\frac{3}{2}}} + \frac{c \| q_0 \|^2}{ \omega^{\frac{3}{2}}},
	\edm
	and
	\bdm
		2 \Re \braket{v_{\al,\omega}}{w_{\al,\omega}} = \frac{2 |Q_0|^2}{(2E_0)^{\frac{3}{2}}} +  \frac{Q^*_0}{\sqrt{\omega} (2E_0)^{\frac{3}{2}}} \int_{\erre} dx \: |x| \Psi_0^{(1)}(x) \: \tilde{q}_0(x) \exp \lf\{ - \sqrt{\frac{2E_0}{\omega}} |x| \ri\} \geqslant  \frac{2 |Q_0|^2}{(2E_0)^{\frac{3}{2}}} - \frac{c \| q_0 \|^2}{ \omega^{\frac{3}{2}}}
	\edm 
	so that $ \lf\| v_{\al,\omega} - w_{\al,\omega} \ri\|^2 \leqslant c \| q_0 \|^2 / \omega^{\frac{3}{2}} $ and \eqref{asygstate} is proven.
	\newline
	Let us now consider the charge $ q_0 $: The asymptotic behavior of the operator $ K^{E_0}_{\omega} $, as $ \omega \to \infty $, is given by the following estimate
	\beq
	\label{projector}
		\lf\| K^{E_0}_{\omega} - \frac{1}{\sqrt{2E_0}} \ket{\Psi^{(\omega)}_0} \bra{\Psi^{(\omega)}_0}  \ri\| = \mathcal{O} \lf( \omega^{-\half} \ri),
	\eeq
	which easily follows from \eqref{carlone2} and the simple inequality
	\bdm
		\lf\| \lf( 2\Hr + 2E_0 - 1 \ri)^{-\frac{1}{2}} -  \frac{1}{\sqrt{2E_0}} \ket{\Psi^{(\omega)}_0} \bra{\Psi^{(\omega)}_0}  \ri\| = \mathcal{O} \lf( \omega^{-\half} \ri).
	\edm
	Therefore by projecting the homogeneous equation \eqref{statleg} onto $ \xi $, we get 
	\bdm
		\| \xi \|^2 - \frac{c |\al| \: \| \xi \| \: \|q_0\|}{\sqrt{\omega}} \leqslant 0
	\edm
	because of \eqref{projector}, which in turn implies $ \| q_0 \|^2 = |Q_0|^2 + \mathcal{O}\lf(\omega^{-1}\ri) $ and $ \| \xi \| = \mathcal{O}\lf(\omega^{-1}\ri) $.
	\newline
	In order to derive from \eqref{asygstate} the $L^2-$convergence of the ground state to $ w_{\al,\omega} $, we need then to bound $ \| q_0 \| $ uniformly in $ \omega $, but can be done by exploiting the $L^2-$normalization of $ u_{\al,\omega} $: \eqref{wnorm},\eqref{asygstate} and the estimate for $ \| \xi \| $ yield
	\bdm
		\frac{|Q_0|}{(2E_0)^{\frac{3}{4}}} = \| w_{\al,\omega} \| \leqslant \| u_{\al,\omega} \| + \frac{c \|q_0\|}{\omega^{\frac{3}{4}}} = 1 + \frac{c \|q_0\|}{\omega^{\frac{3}{4}}} \leqslant 1 + \frac{c |Q_0|}{\omega^{\frac{3}{4}}} + \mathcal{O}\lf(\omega^{-\frac{5}{4}}\ri)  
	\edm
	which together with the reverse inequality implies that $ |Q_0| = 1 + o(1) $ and $ \| q_0 \| = 1 + o(1) $. 
	\newline
	Hence the integral operator with kernel
	\bdm
		\rho^{(w)}_{\al,\omega} (x;\xp) \equiv \int_{\erre} dy \: w^*_{\al,\omega}(x,y) w_{\al,\omega}(\xp,y) = \frac{|Q_0|^2}{2E_0} \exp \lf\{ - \sqrt{2E_0} (|x|+|\xp|) \ri\}
	\edm
	converges in trace class norm to $ \ket{\xi_{\al}} \bra{\xi_{\al}} $, since it is a projector onto a vector which converges in $L^2-$norm to $ \xi_{\al} $ ($ |Q_0| \to  1 $ and $ E_0 \to \al^2/2 $, as $ \omega \to \infty $). On the other hand estimate \eqref{asygstate} gives $ \| \rho_{\al,\omega} - \rho^{(w)}_{\al,\omega} \| = o(1) $ and the convergence of the operators in the norm topology implies weak convergence in $ \mathscr{B}_1(L^2(\erre)) $, but, since $ \mathrm{Tr}(\rho_{\al,\omega}) = \mathrm{Tr}(\ket{\xi_{\al}} \bra{\xi_{\al}}) = 1 $, the convergence is actually in trace class norm. 
\end{dem}

\nt
Now we give some partial results on the positive spectrum of \eqref{ham}.
\spazietto

\begin{teo}[Positive Spectrum of $H_{\al}^{\omega}$]
	\mbox{}	\\
	The essential spectrum of $ H_{\al}^{\omega} $  is equal to $ [0,+\infty) $ and the wave operators $\Omega_{\pm}(H_{\al}^{\omega},H_0^{\omega})$ exist and are complete.
\end{teo}

\begin{dem}
	It is sufficient to prove that $(H_{\al}^{\omega} +\la )^{-1} - (H_{0}^{\omega} +\la )^{-1} $ is a trace class operator
	for some $\la>0$ and $ \omega = 1 $, then the thesis follows from Weyl's Theorem (see Theorem XIII.14 in \cite{rs4}) and Kuroda-Birman Theorem (see Theorem XI.9 in \cite{rs3}). 

	\nt
	Let us introduce the operator $Q^{\la} \equiv (I +\al K^{\la}_{1} )^{-1}$. 
	For $\la >\la_0$, $Q^{\la}$ is bounded and positive (see Proposition \eqref{opek}); 
	the resolvent equation \eqref{res1d} can be cast in the following form
	\begin{equation}
		(H_{\al} +\la)^{-1} - G^{\la} = \Gt \, Q^{\la} \, \Gta.
		\label{diffres}
	\end{equation}
	It is immediate to notice that the r.h.s. of \eqref{diffres} is a positive operator; 
	using Cauchy-Schwartz inequality and the boundedness of $Q^{\la}$, on can prove that 
	$\Gt \, Q^{\la} \, \Gta(x,y;\xp,\yp)$ is a continuous bounded function and that 
	\begin{equation}
		\lf|\Gt \, Q^{\la} \, \Gta (x,y;\xp,\yp)\ri| \leqslant c
		\lf[\int_0^1 d\nu \frac{\nu^{\frac{\la}{\omega} -1}}{\sqrt{ 1 -\nu^2 + \ln  \frac{1}{\nu}}} \ri]^2.
	\end{equation}
	
	\nt
	Then using the Lemma at pg. 65 in \cite{rs3}, together again with Cauchy-Schwartz inequality
	and the boundedness of $Q^{\la}$, one has
	\begin{multline*}
		\int_{\err} dx  dy\, \Gt \, Q^{\la} \, \Gta (x,y;x,y) \leqslant  c
		\int_{\err} dx  dy \lf[ \int_{\erre} dy' \lf| \Gt ( x,y; y',y') \ri|^2 \ri]^{1/2}
		 \lf[ \int_{\erre} dy' \lf| \Gta (y',y'; x,y )\ri|^2 \ri]^{1/2} \leqslant \\
		\leqslant c\int_{\err} dx  dy  dy' \lf| \Gt ( x,y; y',y') \ri|^2 \leqslant c
	\end{multline*}
\end{dem}

\section{The Two Dimensional Case}

\subsection{Preliminary Results}

\nt
In order to rigorously define the operator \eqref{ham}, we use again the theory of quadratic forms. We refer to \cite{me} for proofs and the heuristic derivation of the quadratic form associated with \eqref{ham}. 
\spazietto

\begin{dfn}[Quadratic Form $ \Fa $]
	\label{defforma2d}
	\mbox{}	\\
	The quadratic form $\lf( \Fa, {\mathscr D}( \Fa) \ri)$ is defined as follows
	\begin{equation}
		\label{dforma2d}
		{\mathscr D } ( F_{\al}^{\omega} ) = \lf\{ u \in L^2(\erre^4) \: \big| \:  \exists q \in {\mathscr D}(\Phi^{\la,\omega}_{\al}) \,,\, \varphi^{\la} \equiv  u - \Gt_{\omega} q \in{\mathscr{D}}({F}^{\omega}_{0}) \ri\},
	\end{equation}
	\begin{equation}
		\label{forma2d}
		{F}_{\al}^{\omega}[u]  \equiv  {\mathcal{F}}^{\la,{\omega}}[u] + {\Phi}^{\la}_{\al,\omega}[u],
	\end{equation}
	where \( \lambda > 0 \) is a positive parameter and
	\begin{equation}
		\label{Fforma2d}
		{\mathcal F}^{\la,{\omega}}[u] \equiv \int_{\erre^4} d\vec{x} d\vec{y} \lf\{ \frac{ 1}{2} \absq{ \nabla_{\vec{x}} \fhi^{\la}} + \frac{ 1}{2} \absq{ \nabla_{\vec{y}} \fhi^{\la} }  + \la  \absq{  \fhi^{\la} } -  \la \absq{u} - \omega \absq{u} + \frac{\omega^2 y^2}{2}  \absq{\fhi^{\la} } \ri\},
	\end{equation}
	\begin{displaymath}
		{\mathscr D}(\Phi^{\la}_{\al,\omega})= \lf\{ q \in L^2(\erre^2) \: \big| \: \Phi^{\la,\omega}_{\al}[q] < +\infty \ri\},
	\end{displaymath}
	\begin{equation}
		\label{Phiforma2d}
		\Phi^{\la}_{\al,\omega}[q] \equiv \int_{\erre^2} d\vec{x} \: \left( \al + a^{\la}_{\omega}(x) \right) \absq{q(\vec{x})} + \half \int_{\erre^4} d\vec{x} d\xvp \:  G^{\la}_{\omega} (\vec{x},\vec{x} ; \xvp,\xvp) \absq{ q(\vec{x}) - q(\xvp) },
	\end{equation}
	\beq
		\label{alambda2d}
		a^{\la}_{\omega}(x) \equiv \frac{1}{4\pi} \lf\{ C + \int_{0}^{1} d\!\nu \: \frac{1}{(1- \nu) } \lf[ 1 -  \frac{4 \nu^{ \la/\omega -1 } \lf( 1-\nu \ri)}{\lf( 1+\nu^2 \ri) \ln \frac{1}{\nu} + 1 - \nu^2} \exp \lf( - \frac{\lf( 1 - \nu^2 \ri) \ln \frac{1}{\nu} + 2(1 - \nu)^2}{2 \lf[ \lf( 1 + \nu^2 \ri) \ln \frac{1}{\nu} + 1 - \nu^2 \ri]} \omega x^2  \ri) \ri] \ri\},
	\eeq
	\begin{equation}
		\label{constant 2d}
		C \equiv -\lf( \int_0^{1} d\!\nu \: \frac{e^{-\frac{1}{\nu}}}{\nu} + \int_1^{\infty} d\! \nu \: \frac{e^{-\frac{1}{\nu} }}{\nu^2} \ln \frac{1}{\nu} \ri).
	\end{equation}
\end{dfn}

\nt
Note that the decomposition $ u = \varphi^{\la} + \Gt_{\omega} q $ is well defined and unique (for fixed $ \lambda $), $ \Gt_{\omega} q \in L^2(\erre^4) $ for any $ q \in L^2(\erre^2) $ and the quadratic form \eqref{forma2d} is independent of the parameter $ \lambda $ (see \cite{me}); as a matter of fact $\la$ plays here the role of a free parameter and its value will be chosen later.
\spazietto
\begin{teo}[Closure of the Form $ \Fa $]
	\mbox{}	\\
	The quadratic form $\lf( \Fa,   {\mathscr D}( \Fa) \ri)$ is closed and bounded below on the domain \eqref{dforma2d} for any $ \omega \geqslant 0 $.
\end{teo}

\nt
Let us denote by $\galao$ the positive self adjoint operator on $L^2(\erre^{2})$ associated with the quadratic form $\Phi^{\la}_{\al,\omega}$, i.e. 
\beq
	\label{gala}
	\bra{q} \galao \ket{q} \equiv \Phi^{\la}_{\al, \omega}[q] - \al \| q \|^2,
\eeq
and by $ H_{\al}^{\omega} $ the hamiltonian defined by $ F_{\al}^{\omega} $. Then we have
\spazietto

\begin{teo}[Operator $ H_{\al}^{\omega} $]
	\mbox{}	\\
	The domain and the action of $H_\al^{\omega}$ are the following
	\begin{equation}
	\label{dhamiltonian2d}
		{\mathscr D} ( H_\al^{\omega} )= \lf\{ u \in L^{2}(\erre^{4} ) \: \big| \: u=\fhi^{\la}+ \Gt_{\omega} q, \, \fhi^{\la} \in{\mathscr D} ( H_0^{\omega} ), q\in {\mathscr D} (\galao ), \lf( \al  + \galao \:\ri) q  = \I \fhi^{\la} \ri\},
	\end{equation}
	\begin{equation}
	\label{hamiltonian2d}
		( H_\al^{\omega} +\la ) u = ( H_0^{\omega} +\la )\fhi^{\la},
	\end{equation}
	and the resolvent of $H_{\al}^{\omega}$ can be represented as
	\begin{equation}
	\label{resolvent2d}
		 \lf( H_{\al}^{\omega} + \la \ri)^{-1} f  =  G^{\la}_{\omega} f   + \Gt_{\omega} q_f ,
	\end{equation}
	where, for any $ f \in L^2(\erre^4) $, $q_f$ is a solution of
	\begin{equation}
	\label{eqresolvent2d}
		\lf( \al  + \galao \: \ri) q_f  = \I G^{\la}_{\omega} f .
	\end{equation}
\end{teo}

\nt
As in the one dimensional case, one can recognize in \eqref{hamiltonian2d} a self adjoint extension of the symmetric operator
$\widetilde{H_0}$ defined by
\begin{equation*}
	{\mathscr D}(\widetilde{H_0} )= \lf\{ u \in C_0^{\infty}(\erre^{4} \setminus  \piano ) \ri\}
\end{equation*}
\begin{equation*}
	\widetilde{H_0}u = \lf[ - \half \Delta_{\vec{x}} -\half \Delta_{\vec{y}} + \frac{\omega y^{2}}{2}   - \omega \ri] u
\end{equation*}
so that \( H_{\al}^{\omega} \) is a singular perturbation of \( H_0^{\omega} \) supported on $ \piano $.
\newline
Note that the unperturbed Hamiltonian, corresponding to the case of no interaction
between the particle and the harmonic oscillator, belongs to the family $ H_{\al}^{\omega} $ and is given by $ \al = + \infty $. This fact, which could seem surprising, if one considers the formal Hamiltonian \eqref{ham}, is due to the renormalization procedure required to give a rigorous meaning to such a formal expression. Therefore we stress
that in the two and three-dimensional case $\al$ does not play the role of coupling constant of the
system.
Also in the two dimensional case, the one parameter family of extensions considered have local boundary conditions (see \cite{me}).

\subsection{Spectral Analysis}

\nt
In order to study the spectrum of $ H_{\al}^{\omega} $, we first need to state some spectral properties of the operator $ \galao $:
\spazietto

\begin{prop}[Spectral Analysis of $ \galao $]
	\label{gala spectrum}
	\mbox{}	\\
	For $\omega >0$ the domain $ {\mathscr D}(\Phi^{\la}_{\al,\omega} ) $ can be characterized in the following way:
	\beq
	\label{Gamma domain} 
		{\mathscr D}(\Phi^{\la}_{\al,\omega} ) = 
		\left\{q\in L^2(\erre^2)   \: \Big| \: q \in \hlog\, , \,\hat{q} \in \hlog \right\}.
	\eeq   	
		\newline
	\n
	On this domain $\Phi^{\la}_{\al,\omega}$ is closed and defines a self-adjoint operator $\galao$. For any $ \la >0 $ the spectrum, $ \sigma(\galao) $, is purely discrete, i.e., $ \sigma(\galao) = \sigma_{\mathrm{pp}}(\galao) $. 
	\newline
	Let $ \gamma_n(\la) $, $ n \in \enne $, be the eigenvalues of $ \gala_1 $ arranged in an increasing order $( \lim_{n \rightarrow \infty} \ga_n(\lambda) = + \infty )$. 
	For every $ n \in \enne $, $ \ga_n(\la) $ is a non-decreasing function of $ \la $. Furthermore $ \lim_{\la \rightarrow 0} \ga_0(\la) = -\infty $ and the other eigenvalues remain bounded below, i.e., for any $ \la \geqslant 0 $, there exists a finite constant $ c $  such that $ \ga_n(\la) \geqslant -c $, for any $ n \in \enne$, $ n > 0 $.
\end{prop} 

\begin{dem}
	For the sake of simplicity we fix $ \omega = 1 $ from the outset and omit the dependence on $ \omega $ in the notation.
	\newline
	The self-adjointness of  $ \gala $ immediately follows from the properties of the quadratic form $ \Phi_{\al}^{\la} $ (see \cite{me}). Note that  $ \gala $ can be written in the following way	
	\bdm
		 \gala = a^{\la} + \gala_0,
	\edm
	where $ a^{\la} $ is the multiplication operator for the unbounded function \eqref{alambda2d} and $ \gala_0 $ is the self-adjoint operator associated with the positive quadratic form
	\beq	
		\Phi_0^{\la} [q] \equiv \half \int_{\erre^2} d\xv d\xvp \:  G^{\la}(\xv,\xv ; \xvp,\xvp) \absq{ q(\xv) - q(\xvp) }.
	\label{tomoe}
	\eeq
	Since $ \Phi_0^{\la} $ is positive and $ a^{\la}(x) $ is bounded below but not 
	above,  $ \gala $ is an unbounded operator. 
	Notice that $ a^{\la}(x) $ is a monotone increasing function of $x$ and $a^{\la}(x)  \simeq c \log x$ 
	for $x \ten \infty$; furthermore $ a^{\la}(x)  \geqslant  a^{\la}(0) $, $a^{\la}(0)$ is a monotone 
	increasing function of $\la$ and $a^{\la}(0) \simeq c\log \la$ for $\la \ten \infty$. Hence 
	the following lower bound
	\begin{equation}
	\Phi_{\al}^{\la} [q] \geqslant 
	\int_{\erre^2} d\vec{x} \: \left( \al + a^{\la}(x) \right) \absq{q(\vec{x})} \geqslant
	\left( \al + a^{\la}(0) \right) \| q \|^2 
	\label{inferiore}
	\end{equation}
	proves that for any
	$\al \in \erre$ there exists $\la_0 $ such that for $\la > \la_0$ the quadratic form $\Phi_{\al}^{\la}$ is positive; 
	for such $\la$ the operator $\gala$ is invertible and its inverse is a bounded operator.
	
	\n
	We shall prove now that $ \sigma(\gala) $ is purely discrete for any $ \la $: By Theorem XIII.64 in \cite{rs4}
	it suffices to prove that
	\bdm
		\mathscr{D}_{\eta} \equiv \lf\{ q \in \mathscr{D}(\Phi_{\al}^{\la}) \: \big| \: \Phi_{\al}^{\la}[q] \leqslant \eta \ri\} 
	\edm
	is a compact subset of $ L^2(\erre^2) $ for any positive $ \eta $; this will proved using
	the Rellich's criterion (Theorem XIII.65 in \cite{rs4}). 
	The positivity of $\gala_0 $ implies that, if $ \Phi_{\al}^{\la}[q] \leq \eta $, then 
	\bdm
		\int_{\erre^2} d\vec{x} \: a^{\la}(x) \absq{q(\vec{x})} \leqslant \eta.
	\edm
	Moreover, applying the Fourier transform, $ \Phi_{\al}^{\la} $ can be rewritten in the following equivalent form 
	\beq
	\label{formafourier2d}
		\Phi_{\al}^{\la}[q] = \int_{\erre^2} d\vec{k} \: \left( \al + \tilde{a}^{\la}(k) \right) \absq{\hat{q}(\vec{k})} + \half \int_{\erre^4} d\vec{k} d\vec{k}^{\prime} \:  \tilde{G}^{\la}(\vec{k} ; \vec{k}^{\prime}) \absq{ \hat{q}(\vec{k}) - \hat{q}(\vec{k}^{\prime}) },
	\eeq 
	where 
	\beq
	\label{afourier2d}
		\tilde{a}^{\la}(k) \equiv \frac{1}{4\pi} \lf\{ C + \int_0^1 d\nu \: \frac{1}{1-\nu} \lf[1 - \frac{4 \nu^{\la-1} (1-\nu)}{(1+\nu^2)\ln \frac{1}{\nu} + 1 - \nu^2} \exp \lf( - \frac{(1-\nu^2)\ln \frac{1}{\nu}}{2 \lf[ (1+\nu^2) \ln \frac{1}{\nu} + 1 - \nu^2 \ri]} k^2 \ri) \ri] \ri\},
	\eeq
	\begin{multline}
	\label{greenfourier2d}
			 \tilde{G}^{\la}(\vec{k} ; \vec{k}^{\prime}) \equiv \frac{1}{2\pi^2} \int_0^1 d\nu \: \frac{\nu^{\la-1}}{(1-\nu^2) \ln \frac{1}{\nu} + 2(1-\nu)^2} \cdot	\\
		\cdot \exp \lf\{ - \frac{\lf[ (1+\nu^2) \ln \frac{1}{\nu} + 1 - \nu^2 \ri] \lf( k^2 + {k^{\prime}}^2 \ri)}{2 \lf[(1-\nu^2) \ln \frac{1}{\nu} + 2(1-\nu)^2 \ri]} - \frac{\lf[ 1 - \nu^2 + 2\nu \ln \frac{1}{\nu} \ri]  \vec{k} \cdot \vec{k}^{\prime}}{(1 - \nu^2) \ln \frac{1}{\nu} + 2(1 -\nu)^2} \ri\}.
	\end{multline}
In order to prove \eqref{formafourier2d}, it is convenient to introduce a regularized quadratic form
$\Phi^{\la}_{\al ,\de} $ obtained by restricting 
the integration domain in $ \nu $ to the set $ [0,1-\de] $, for some $ 0 < \de < 1 $.

\n
It is straightforward to notice that $\Phi^{\la}_{\al, \de} $ is a bounded form and that for every $q \in L^2(\erre^2) $, 
$\Phi^{\la}_{\al, \de}[q] $ is a monotone function of $\de$; therefore for $q\in \mathscr{D}( \Phi_{\al}^{\la}) $ we have 
\begin{equation}
\lim_{\de \ten 0} \Phi^{\la}_{\al, \de}[q] = \Phi_{\al}^{\la} [q].
\label{kenshin}
\end{equation}
On the other hand, due to the regularization, with straightforward calculations one can prove that
\begin{multline}
		\Phi_{\al, \de}^{\la}[q] \equiv \lf\{ \al +  \frac{1}{4\pi} \lf[ C + \int_0^{1-\de} \frac{d\nu}{1-\nu} \ri] \ri\} \: \| q \|^2 - \int_{\erre^4} d\vec{x} d\xvp \:  G_{\de}^{\la}(\vec{x},\vec{x} ; \xvp,\xvp) \: q^*(\vec{x}) q(\xvp) = \\=
		 \lf\{ \al +  \frac{1}{4\pi} \lf[ C + \int_0^{1-\de} \frac{d\nu}{1-\nu} \ri] \ri\} \: \| \hat{q} \|^2 - \int_{\erre^4} d\vec{k} d\vec{k}^{\prime} \:  \tilde{G}_{\de}^{\la}(\vec{k} ; \vec{k}^{\prime}) \: 
		 \hat{q}^*(\vec{k}) \hat{q}(\vec{k}^{\prime}),
		 \label{mejii}
\end{multline}
which can be rewritten in the following way:
	\begin{equation}
		\Phi_{\al,\de}^{\la}[q] = \int_{\erre^2} d\vec{k} \: \left( \al + \tilde{a}_{\de}^{\la}(k) \right) \absq{\hat{q}(\vec{k})} + \half \int_{\erre^4} d\vec{k} d\vec{k}^{\prime} \:  \tilde{G}_{\de}^{\la}(\vec{k} ; \vec{k}^{\prime}) \absq{ \hat{q}(\vec{k}) - \hat{q}(\vec{k}^{\prime}) },
	\label{ruroni}
	\end{equation}
where $\tilde{a}_{\de}$ and $\tilde{G}_{\de}^{\la}$ are the regularization of \eqref{afourier2d} and 
\eqref{greenfourier2d}. Notice that \eqref{mejii} shows how $ \Phi_{\al}^{\la}$ can be obtained by a renormalization of the
formal quantity $\bra{q} \Gt \ket{q}$.

\n
Due to the monotonicity in $\de$, we can take the limit $\de \ten 0$ of \eqref{ruroni} and, by \eqref{kenshin},
we obtain \eqref{formafourier2d}. It is immediate to notice that \eqref{formafourier2d} has the same
structure as \eqref{Phiforma2d} and in particular, if  $ \Phi_{\al}^{\la}[q] \leq \gamma $, then
	\bdm
		\int_{\erre^2} d\vec{k} \: \tilde{a}^{\la}(k) \absq{\hat{q}(\vec{k})} \leqslant \eta.
	\edm
	The function $ \tilde{a}^{\la}(k) $ has the same properties of $ a^{\la}(x) $, namely 
	it is a monotone function of $k$, $\hat{a}^{\la}(k) \simeq  c \log k$ for $k\to \infty$ and 
	$\hat{a}^{\la}(0) \simeq  c \log \la$ for $\la \to 0$. 
	Hence Rellich's criterion guarantees that $ \mathscr{D}_{\eta} $ is a compact subset of $ L^2(\erre^2) $ 
	and therefore $\gala$ has only pure point spectrum. 
	
	\n
	Notice that also the following bound holds,
	\begin{equation}
	\Phi^{\la}_0[q] \leqslant c \| q\|^2_{ {\mathcal H}^{ \text{log} }(\erre^2)  }.
	\label{bakumatsu}
	\end{equation}
	Indeed using the following inequality 
	\begin{equation*}
	G^{\la}(\xv,\xv ; \xvp,\xvp) \leqslant c 	
	 \int_0^1 d\nu \: \frac{\nu^{ \la - 1 }}{\lf( 1- \nu^2 \ri)  \ln \frac{1}{\nu} } 
	 \exp \lf\{  -   \frac{1}{ 2\ln {\frac{1}{\nu}  }  } (\xv-\xvp)^2 - \frac{ \nu}{ 1 - \nu^2}  ( \xv - \xvp )^2 \ri\}
\end{equation*}
in \eqref{tomoe} and taking the Fourier transform, we have
\begin{equation}
\Phi^{\la}_0[q] \leqslant c \int_{\erre^2} d\vec{k}
 \int_0^1 d\nu \:	\frac{\nu^{ \la - 1 }}{ 2\nu \ln \frac{1}{\nu} + 1- \nu^2 } 
\lf\{ 1-  \exp \lf[  - \frac{ 2(1- \nu^2) \ln \frac{1}{\nu} }{ 4( 2\nu \ln \frac{1}{\nu} + 1- \nu^2) } k^2 \ri] \ri\} \:  \lf|\hat{q}(k) \ri|^2 ,
 \end{equation}
and, since
\begin{equation*}
  0\leqslant \int_0^1 d\nu \:	\frac{\nu^{ \la - 1 }}{ 2\nu \ln \frac{1}{\nu} + 1- \nu^2 } 
\lf\{1-  \exp \lf[  - \frac{ 2(1- \nu^2) \ln \frac{1}{\nu} }{ 4( 2\nu \ln \frac{1}{\nu} + 1- \nu^2) } k^2 \ri] \ri\}
 \leqslant c \log (1 + \Mk), 
 \end{equation*}
 \eqref{bakumatsu} is proven. Therefore, taking into account the behavior of $a^{\la}$, 
 \eqref{bakumatsu} implies that exist $\la_0 > 0$ such that for $\la > \la_0$ we have
 \begin{equation}
 \gala \leqslant c \lf( \log \Mx + \log \Mp + \log \la \ri).
 \label{pigneto}
 \end{equation}

\n
We can prove also a similar lower bound for $\gala$:
Putting together the lower bound \eqref{inferiore} and a corresponding lower bound for  
\eqref{formafourier2d}, we obtain:
	\begin{equation}
	\bra{q} \gala \ket{q}  \geqslant 
	\half \int_{\erre^2} d\vec{x} \:  a^{\la}(x) \ \absq{q(\vec{x})} +
	\half \int_{\erre^2} d\vec{k} \:  \hat{a}^{\la}(k) \ \absq{\hat{q}(\vec{k})} 
	\geqslant
	\lf[ \frac{a^{\la}(0)}{2}  + \frac{\hat{a}^{\la}(0)}{2} \ri] \| q \|^2
	\label{lowerbene}
	\end{equation}
and in particular \eqref{Gamma domain} holds true, due to \eqref{pigneto} and \eqref{lowerbene}.

\n
	The monotonicity in ${\la}$ of the eigenvalues $\ga_n (\la)$ follows from monotonicity of  $  \Phi_{\al}^{\la}[q] $ with respect to $ \la $. This can be easily seen by observing that the regularized expression 
	\eqref{mejii} is a non-decreasing function of $ \la $, as it must be its limit as $ \de \to 0 $.
	\newline
	In order to analyze the asymptotics for $ \la \rightarrow 0 $ of $ \ga_0(\la) $, we shall show that there exists a function $ q $ belonging to $ \mathscr{D}(\gala) $, such that $ \lim_{\la \rightarrow 0} \bra{q} \gala \ket{q} = - \infty $. The result is then a consequence of the Min-Max Theorem. Indeed taking the ground state of the 2d harmonic oscillator $ \Psi^{(1)}_0(\xv) $, one has
	\begin{multline}
	\label{kyoto}
		\lim_{\la \to 0} \bra{ \Psi^{(1)}_0} \gala \ket{ \Psi^{(1)}_0} = \lim_{\la \to 0} \frac{1}{4 \pi} \lf\{ C + \int_0^1 d\nu \: \frac{1}{1-\nu} \lf[ 1 - \frac{8 \nu^{\la-1} (1 - \nu)}{(3+\nu^2) \ln \frac{1}{\nu} + 4(1-\nu)} \ri] \ri\} \leqslant \\
		c_1 -  c_2 \lim_{\la \to 0} \int_0^{\half} d\nu \: \frac{\nu^{\la-1}}{1+\ln \frac{1}{\nu}} = - \infty.
	\end{multline}
	
	\n
The boundedness from below of the other eigenvalues can be proved by showing that the quadratic form remains bounded as $ \la \to 0 $, if $ q $ is orthogonal to the above function $ \Psi^{(1)}_0(\xv) $: Let $ q^{\perp}(\xv) $ be a $L^2-$normalized function\footnote{For instance one can take $ q^{\perp} = \Psi_n^{(1)} $, $ n > 0 $.} in $ \mathscr{D}( \Phi_{\al}^{\la}) $ such that $ \braket{ \Psi^{(1)}_0(\xv) }{q^{\perp}} = 0 $. From the expression of the quadratic form \eqref{Phiforma2d}, it is clear that we can restrict the integration in $ \nu $ in \eqref{alambda2d} and in $ G^{\lambda} $ to the interval $ [0,1/e] $, because the remainder is uniformly bounded in $ \la $, so that, acting as in \eqref{mejii}, we have
	\bdm
		 \bra{q^{\perp}} \gala \ket{q^{\perp}} \geqslant \frac{1}{4\pi} \lf[ C + \int_0^{\frac{1}{e}} d\nu \: \frac{1}{1-\nu} \ri] \lf\| q^{\perp} \ri\|^2 - \int_{\erre^4} d\vec{x} d\xvp \:  G_{1/e}^{\la}(\vec{x},\vec{x} ; \xvp,\xvp) \: \lf( q^{\perp}(\vec{x}) \ri)^* q^{\perp}(\xvp) 
	\edm
	\bdm	
		\int_{\erre^4} d\vec{x} d\xvp \:  G_{1/e}^{\la}(\vec{x},\vec{x} ; \xvp,\xvp) \: \lf( q^{\perp}(\vec{x}) \ri)^* q^{\perp}(\xvp) \leqslant \frac{1}{2\pi^2} \int_0^{\frac{1}{e}} d\nu \: \frac{\nu^{\la-1}}{1-\nu^2} \bra{q^{\perp}} k_{\nu} \ket{q^{\perp}}  \equiv  \bra{q^{\perp}} K^{\la} \ket{q^{\perp}},
	\edm
	where $ k_{\nu} $ is the integral operator whose kernel is the two-dimensional analogous of \eqref{pastiera2}. Moreover
	\beq
	\label{diffresharm}
		 \bra{q^{\perp}} K^{\la} \ket{q^{\perp}} \leqslant  \frac{1}{2\pi} \bra{q^{\perp}} (\Hro + \la - 1)^{-1} \ket{q^{\perp}} + \frac{1}{2\pi^2} \int_0^{\frac{1}{e}} d\nu \: \frac{\nu^{\la-1}}{1-\nu^2} \bra{q^{\perp}} k_{\nu} - \bar{k}_{\nu} \ket{q^{\perp}},
	\eeq
	$  \bar{k}_{\nu} $ denoting the integral operator with kernel
	\beq
		 \bar{k}_{\nu}(x,\xp) \equiv \exp \lf\{ -\frac{1}{2} \frac{1-\nu}{1+\nu} (x^2 + {\xp}^2) - \frac{\nu (\xv-\xvp)^2}{1-\nu^2} \ri\}.
	\eeq
	The last term in \eqref{diffresharm} can be estimated as follows
	\begin{multline*}
		\frac{1}{2\pi^2} \int_0^{\frac{1}{e}} d\nu \: \frac{\nu^{\la-1}}{1-\nu^2} \bra{q^{\perp}} k_{\nu} - \bar{k}_{\nu} \ket{q^{\perp}} \leqslant \frac{1}{4\pi^2} \int_0^{\frac{1}{e}} d\nu \: \frac{\nu^{\la-1}}{(1-\nu^2) \ln\frac{1}{\nu}} \int_{\erre^4} d\vec{x} d\xvp \: (\xv-\xvp)^2 \bar{k}_{\nu}(\xv;\xvp) \lf|q^{\perp}(\xv)\ri| \lf|q^{\perp}(\xvp)\ri| \leqslant	\\
	\frac{1}{\pi^2} \int_0^{1} d\nu \: \frac{\nu^{\la-1}}{1-\nu^2} \int_{\erre^4} d\vec{x} d\xvp \: x^2 \bar{k}_{\nu}(\xv;\xvp) \lf|q^{\perp}(\xv)\ri| \lf|q^{\perp}(\xvp)\ri| \leqslant \frac{1}{\pi} \bra{\lf|q^{\perp}\ri|} \Hro (\Hro + \la - 1)^{-1} \ket{\lf|q^{\perp}\ri|} \leqslant \frac{ \lf\| q^{\perp} \ri\|^2}{\pi}.
	\end{multline*}
	Since, for any $ q^{\perp} $ orthogonal to the ground state of $ \Hro $,  $ \bra{q^{\perp}} (\Hro + \la - 1)^{-1} \ket{q^{\perp}} \leqslant \lf\| q^{\perp} \ri\|^2 $, we thus obtain
	\bdm
		 \bra{q^{\perp}} K^{\la} \ket{q^{\perp}} \leqslant \frac{3 }{2\pi} \lf\| q^{\perp} \ri\|^2
	\edm
	and the boundedness from below of the operator $ \gala $ on the subspace of functions orthogonal to $ \Psi^{(1)}_0 $.
\end{dem}	
  
\nt
The spectral properties of the operator  $ \galao $ allow us to give a complete characterization of the discrete spectrum of $ H_{\al}^{\omega}$:
\spazietto

\begin{teo}[Negative Spectrum of $H_{\al}^{\omega}$]
	\label{discrete spectrum 2d}
	\mbox{}	\\
	For any $ \al \in \erre $ and $\omega \in \erre^+$ the discrete spectrum $ \sigma_{\mathrm{pp}}(H_{\al}^{\omega}) $ of $ H_{\al}^{\omega} $ is not empty and it contains a number $ N_{\omega}(\al)\geqslant 1 $ of negative eigenvalues 
$-E_0(\al,\omega) \leqslant -E_1(\al,\omega) \leqslant \ldots\leqslant 0 $, satisfying the scaling 
	\beq 
	\label{scaling2d}
		E_n(\al,\omega) = \omega E_n(\al,1).
	\eeq 
	The corresponding eigenvectors are given by $ u_n = {\mathcal G}^{E_n}_{\omega}  q_n $, where $ q_n $ is a solution of the homogeneous equation $ \al q_n + \Gamma^{E_n}_{\omega} q_n = 0 $. 
	\newline
	Moreover there exists $ \alpha_0 \in \erre $ such that, if $ \al > \al_0 $, $ N_{\omega}(\al) = 1 $ and, for fixed $ \omega $ and $\al \to - \infty$, $\ln N_{\omega}(\al) \geqslant c |\al|$. 	
	\newline
 The ground state energy has the following asymptotic behavior for fixed $ \omega $: $E_0\simeq -c \al^{-1}$ for $\al \ten +\infty$ and $\ln  E_0 \simeq c|\al| $ for $\al \ten - \infty$. 
	
\end{teo}

\begin{dem}
	Following the proof of Theorem \ref{spectral1d}, we get that $ u_E $ is an eigenfunction of $ H_{\al}^{\omega} $ relative to the eigenvalue $ -E $, $E > 0$, only if
	\beq
		u_E = {\mathcal G}^{E}_{\omega} q
	\eeq
	for some $ q \in {\mathscr D(\Phi^{\la}_{\al,\omega})  }  $. On the other hand $ u_E $ belongs to the domain of $ H_{\al}^{\omega} $ and then it must satisfies the boundary condition on $ \piano $, which for a function of this form becomes $ \al q + \gae_{\omega} q = 0 $, or
	\beq
	\label{statleg2d}
		\Phi^{E}_{\al,\omega}[q] = \alpha \lf\| q \ri\|^2  +  \bra{q} \gae_{\omega} \ket{q} = 0. 
	\eeq
	So that there is a one-to-one correspondence between the negative eigenvalues of $ H_{\al}^{\omega} $ and non trivial solutions of the homogeneous equation above. In other words $ - E $ is an eigenvalue of $ H_{\al} $, if and only if $ 0 $ is an eigenvalue of $ \al + \gae_{\omega} $. Note that, by scaling, $ q $ solves \eqref{statleg2d}, if and only if $ \tilde{q}(\xv) \equiv \omega^{-\frac{1}{2}} q(\xv/\sqrt{\omega}) $ is a solution of the homogeneous equation
	\bdm
		\al \tilde{q} + \Gamma^{E/\omega}_1 \tilde{q} = 0,
	\edm
	which implies \eqref{scaling2d}. 
	\newline
	The other results are simple consequences of Proposition \ref{gala spectrum}. In particular in order to complete the asymptotic analysis for $\al \ten +\infty$ it is sufficient to notice 
that in facts \eqref{inferiore} and \eqref{kyoto} imply that $- c_1 \la^{-1}\leqslant 
\ga_0(\la) \leqslant - c_2 \la^{-1}$ as $\la \ten 0$, due to the asymptotic behavior of $a^{\la}_{\omega}(0)$ and
$\hat{a}^{\la}_{\omega}(0)$ in such a limit; this is sufficient to conclude that $E_0 = \mathcal{O}( \al^{-1})$ for $\al \ten +\infty$.

\n
The previous argument can be repeated for $\la \ten - \infty$ and gives that $\ga_0(\la) \geqslant c \ln \la  $ for 
$\la \ten + \infty$, which means $\ln  E_0 = \mathcal{O}(|\al| ) $ for $\al \ten - \infty$.

\n
In order to conclude the proof, it is sufficient to notice that, for fixed $ \omega = 1 $, $N_{1}(\al)$ is bounded below by the cardinality of
$\{ n \in \enne \: | \: \ga_{n} (0) \leqslant - \al \}$; therefore any upper bound on $\ga_{n} (0)$ provides
a lower bound on $N_{1}(\al)$. Due to the monotonicity in $\la$ of $\ga_{n} (\la)$, to \eqref{pigneto} and to the 
straightforward estimate $\lf( \log \Mx + \log \Mp  \ri) \leqslant c \log ( \Hro + 1 )$, we can
use the eigenvalue distribution of the logarithm of the harmonic oscillator to estimate $N_{1}(\al)$
which gives $\ln N_{1}(\al) \geqslant c |\al|$ for $\al \to \infty$.
\end{dem}

\nt
We underline that for $\omega >0$ the interaction is attractive in the sense that there exists at least one bound state irrespective of the sign of $ \al $. This fact is essentially due to the renormalization procedure used to rigorously define the quadratic form in \eqref{forma2d} and to the presence of the harmonic oscillator; this is a common phenomenon in the theory of point interactions (see, e.g., \cite{al} for a similar effect).
\newline
Note also that the different scaling \eqref{scaling2d} in $ \omega $ is due to the scaling properties of the Green function \eqref{gre} (more precisely its restriction to the planes $ \Pi $), i.e., in $ d $ dimensions,
\bdm
	G^{\la}_{\omega} (\xv,\yv;\xvp,\yvp) = \omega^{d-1} G^{\la/\omega}_{1} (\sqrt{\omega}\xv,\sqrt{\omega}\yv;\sqrt{\omega}\xvp,\sqrt{\omega}\yvp).  
\edm
The asymptotics for $ \omega \to 0 $ can be easily derived from \eqref{scaling2d}: 
The spacing between different eigenvalues goes to 0 and in the limit they form a continuum, so that
no bound state survives in the limit. On the opposite all the eigenvalues 
corresponding to excited states diverge as $ \omega \to \infty $. 

\n
Before giving a partial characterization of the positive spectrum, let us prove a
technical lemma.
\spazietto

\begin{lem}
Let us consider the following operator 
$T_{\omega}^k \equiv \Gta_{\omega} ( G^{\la}_{\omega} )^k \Gt_{\omega} :L^2 (\erre^2)\ten L^2(\erre^2) $, $k\in \enne$. Then, if $ \la > k $,   $T_{\omega}^k \in 
{\mathscr B}_{p} ( L^2 (\erre^2),L^2(\erre^2))$, for any $p>2 (k+1)^{-1} $.
\label{puzzola}
\end{lem}

\begin{dem}
Setting $ \omega = 1 $ for the sake of clarity and omitting the $\omega-$dependence in the notation, we have the identity
\begin{equation}
T^k =  \lf( -\frac{d}{d\la} \ri)^{k+1} \I G^{\la} \I^{\ast},
\label{inu}
\end{equation}
so that, using \eqref{gre}, we get the integral kernel of $T^k$, i.e.,
\begin{equation*}
T^k(\xv;\xvp)= c 
\int_0^1 d\nu \;
\frac{ \nu^{  \la-1  } \lf(   \ln{  \frac{1}{\nu}  }     \ri)^{k }}{ \lf( 1- \nu^2 \ri)  } \exp \lf\{ -\half \nuu  (x^2 + \xpq) -   \frac{(\xv-\xvp)^2}{ 2\ln {\frac{1}{\nu}  }  }  - \frac{ \nu( \xv - \xvp )^2}{ 1 - \nu^2}   \ri\}.
\end{equation*}
Notice that, using the same argument as in the proof of Proposition \ref{opek}, 
we can view $T^k$ as the integral
over the parameter $\nu$ of positive operator valued functions $ t_{\nu} $, i.e.,
\beq
\label{tk}
T^k = c \int_0^1 d\nu \, m_k(\nu) t_{\nu},	
\eeq
\beq
\label{mketnu}
m_k(\nu) =
\frac{ \nu^{  \la -1 }\lf(   \ln{  \frac{1}{\nu}  }     \ri)^{k } }{ \lf( 1- \nu^2 \ri)   },
\qquad
t_{\nu}(\xv;\xvp)= \exp \lf\{ -\half \nuu  (x^2 + \xpq) -   \frac{(\xv-\xvp)^2}{ 2\ln {\frac{1}{\nu}  }  }  - \frac{ \nu( \xv - \xvp )^2}{ 1 - \nu^2}   \ri\}.
\eeq
Applying Schur test to the operator $ t_{\nu} $, one has $ \| t_{\nu} \|_{\infty} = \| t_{\nu} \| \leqslant c (1-\nu) $, whereas a simple calculation yields $ \| t_{\nu} \|_1 = \mathrm{Tr}(t_{\nu}) \leqslant c (1-\nu)^{-1} $. On the other hand H\"{o}lder inequality in Schatten ideals (see \cite{si}) gives 
\beq
\label{tnunorm}
	\| t_{\nu} \|_p \leqslant \|t_{\nu}\|_{1}^{1/p} \| t_{\nu} \|^{1-1/p} \leqslant (1-\nu)^{1-2/p}.
\eeq
It is then straightforward to check that
\begin{equation}
	\| T^k \|_p \leqslant \int_0^1 d\nu \, m_k(\nu) \| t_{\nu} \|_p < + \infty
\end{equation}
for any $ p > 2(k+1)^{-1} $.
\end{dem}

\nt
Now we present some partial results on the continuous spectrum of \eqref{ham} by mean of a 
characterization of the mapping properties of
the resolvent; in the following we shall fix $\la>1$, such that 
$(\Gamma^{\la}_{\omega} + \al )^{-1}$ exists and is bounded. Therefore Eq. \eqref{resolvent2d} can be cast 
in the following form
\begin{equation}
(H_{\al}^{\omega} + \la )^{-1} = G^{\la}_{\omega} - \Gt_{\omega} (\Gamma^{\la}_{\omega} + \al )^{-1} \Gta_{\omega}.
\label{frodo}
\end{equation}
\spazietto

\begin{teo}[Positive Spectrum of $H_{\al}^{\omega}$]
	\mbox{}	\\
	The essential spectrum of $ H_{\al}^{\omega} $  is equal to $ [0,+\infty) $ and the wave operators $\Omega_{\pm}(H_{\al}^{\omega},H_0^{\omega})$ exist and are complete.
\end{teo}

\begin{dem} 
	We shall drop the dependence on $\omega$ for brevity.
	It is sufficient to prove that $(H_{\al} +\la )^{-1} - (H_{0} +\la )^{-1}$ is a compact operator
	and that $[(H_{\al} +\la )^{-1}]^3 - [(H_{0} +\la )^{-1}]^3$ is trace class
	for some $\la>0$, then the thesis follows from Weyl's Theorem (see Theorem XIII.14 in \cite{rs4}) and 
	Corollary 3 of Theorem XI.11 in \cite{rs3}). 
	
	\n
	We first analyze $\Gt ( \gala +\al )^{-1} \Gta $ and prove that it is a compact operator.
	Due to Lemma \ref{puzzola} we have $\Gt \in {\mathscr B}_p(L^2(\erre^2),L^2(\erre^4))$ with $ p > 4 $: By taking $ k = 0 $, one obtains $ \Gta\Gt \in {\mathscr B}_p(L^2(\erre^2),L^2(\erre^2)) $ for $ p > 2 $, i.e., denoting by $ g_n^2 $, $ n \in \enne $, its singular values, $ \{ g_n \} \in \ell_p $ for $ p > 4 $. By a standard argument (see, e.g., the proof of Theorem VI.17 in \cite{rs1}), one can show that $ \{ g_n \} $ are the singular values of $ \Gt $ and the result easily follows. This also implies that $ \Gta \in  {\mathscr B}_p(L^2(\erre^4),L^2(\erre^2)) $, $ p > 4 $, and both operators are compact. Moreover $( \gala +\al )^{-1}$ is a bounded operator and then $ \Gt ( \gala +\al )^{-1} \Gta \in {\mathscr B}_p(L^2(\erre^4),L^2(\erre^4))$ with $p > 2 $, by H\"older inequality, and in particular is a compact operator.
	
	\n
	In order to prove the existence of wave operators and asymptotic completeness, let us expand
	the difference of the resolvent to third power:
	\begin{multline}
[(H_{\al} +\la )^{-1}]^3 - [(H_{0} +\la )^{-1}]^3 =
\Gt (\Gamma^{\la} + \al )^{-1} \Gta \lf( G^{\lambda} \ri)^2 +  
G^{\lambda}\Gt (\Gamma^{\la} + \al )^{-1} \Gta  G^{\lambda} + \\
\lf( G^{\lambda} \ri)^2 \Gt (\Gamma^{\la} + \al )^{-1} \Gta+
\lf(  \Gt (\Gamma^{\la} + \al )^{-1} \Gta       \ri)^2 G^{\lambda}  + \Gt (\Gamma^{\la} + \al )^{-1} \Gta G^{\lambda} \Gt (\Gamma^{\la} + \al )^{-1}\Gta  + \\
 G^{\lambda}  \lf(  \Gt (\Gamma^{\la} + \al )^{-1} \Gta       \ri)^2 +
\lf(  \Gt (\Gamma^{\la} + \al )^{-1} \Gta       \ri)^3.  
\label{chepalle1}
\end{multline}
All the terms on the r.h.s. of \eqref{chepalle1} are trace class operators. Indeed it is sufficient
to use Lemma \ref{puzzola}, with $ k = 2$, and H\"older inequality, as done when studying
$\Gt ( \gala +\al )^{-1} \Gta$. As an example let us consider the first term in the above expression: By Lemma \ref{puzzola}, $ \Gta \lf( G^{\lambda} \ri)^2 \in \mathscr{B}_p(L^2(\erre^4),L^2(\erre^2)) $ for $ p > 4/5 $ (by the same argument applied to $ \Gt $) and thus it is a trace class operator. The claim then follows from boundedness of $ \Gt $ and $ ( \gala +\al )^{-1} $ and H\"{o}lder inequality.
\end{dem}

\section{The Three Dimensional Case}

\subsection{Preliminary Results}

\nt
As in the two-dimensional case, the operator \eqref{ham} can be rigorously defined by means of the theory of quadratic forms (see \cite{me}): 
\spazietto

\begin{dfn}[Quadratic Form $ \Fa $]
	\label{defforma3d}
	\mbox{}	\\
	The quadratic form $\lf( \Fa, \DFa \ri)$ is defined as follows
	\begin{equation}
		\label{dforma3d}
		{\mathscr D } ( F_{\al}^{\omega} ) = \lf\{ u \in L^2(\erre^6) \: \big| \:  \exists q \in {\mathscr D}(\Fia) \,,\, \varphi^{\la} \equiv  u - \Gt_{\omega} q \in{\mathscr{D}}({{F}}^{\omega}_{0}) \ri\},
	\end{equation}
	\begin{equation}
		\label{forma3d}
		{F}_{\al}^{\omega}[u]  \equiv  {\mathcal{F}}^{\la,\omega}[u] + \Fia[u],
	\end{equation}
	where \( \lambda > 0 \) is a positive parameter and
	\begin{equation}
		\label{Fforma3d}
		{\mathcal F}^{\la,\omega}[u] \equiv \int_{\erre^6} d\vec{x} d\vec{y} \lf\{ \frac{ 1}{2} \absq{ \nabla_{\vec{x}} \fhi^{\la}} + \frac{ 1}{2} \absq{ \nabla_{\vec{y}} \fhi^{\la} }  + \la  \absq{  \fhi^{\la} } -  \la \absq{u} - \frac{3\omega}{2} \absq{u} + \frac{\omega^2 y^2}{2} \absq{\fhi^{\la} } \ri\},
	\end{equation}
	\begin{displaymath}
		{\mathscr D}(\Fia)= \lf\{ q \: \big| \: q \in L^2(\erre^3) \, , \, \Fia[q] < +\infty \ri\},
	\end{displaymath}
	\begin{equation}
		\label{Phiforma3d}
		\Fia[q] \equiv \int_{\erre^3} d\vec{x} \: \left( \al + a^{\la}_{\omega}(x) \right) \absq{q(\vec{x})} + \half \int_{\erre^6} d\vec{x} d\xvp \:  G^{\la}_{\omega}(\vec{x},\vec{x} ; \xvp,\xvp) \absq{ q(\vec{x}) - q(\xvp) },
	\end{equation}
	\begin{multline}
		\label{alambda3d}
		a^{\la}_{\omega}(x) \equiv \frac{\sqrt{\omega}}{\left( 4\pi \right)^{\frac{3}{2}}} \lf\{ \frac{1}{2} + \ri. \\
		\lf. \int_{0}^{1} d\!\nu \: \frac{1}{(1- \nu)^{\frac{3}{2}} } \lf[ 1 -  \frac{8 \nu^{ \la/\omega -1 } \lf( 1-\nu \ri)^{\frac{3}{2}}}{\lf[ \lf( 1+\nu^2 \ri) \ln \frac{1}{\nu} + 1 - \nu^2 \ri]^{\frac{3}{2}}} \exp \lf( - \frac{\lf( 1 - \nu^2 \ri) \ln \frac{1}{\nu} + 2(1 - \nu)^2}{2 \lf[ \lf( 1 + \nu^2 \ri) \ln \frac{1}{\nu} + 1 - \nu^2 \ri]} \omega x^2  \ri) \ri] \ri\}. 
	\end{multline}
\end{dfn}

\nt
The well-posedness of the definition above can be shown exactly as in the two-dimensional case. Moreover in the same way one can prove that the form is actually closed and bounded below (see \cite{me} for the proofs):
\spazietto

\begin{teo}[Closure of the Form $ \Fa $]
	\mbox{}	\\
	The quadratic form $\lf( \Fa, \DFa \ri)$ is closed and bounded below on the domain \eqref{dforma3d}.
\end{teo}

\nt
Concerning the self-adjoint operators $ H_{\al}^{\omega} $ and $\galao$ associated with the quadratic forms $ \Fa $ and $ \Fia $ respectively, i.e., 
\beq
	\label{gala3d}
	\bra{q} \gala_{\omega} \ket{q} \equiv \Fia[q] - \al \| q \|^2,
\eeq
we have the following,
\spazietto

\begin{teo}[Operator $ H_{\al}^{\omega} $]
	\mbox{}	\\
	The domain and the action of $H_\al^{\omega}$ are the following
	\begin{equation}
	\label{dhamiltonian3d}
		{\mathscr D} ( H_\al^{\omega} )= \lf\{ u \in L^{2}(\erre^{6} ) \: \big| \: u=\fhi^{\la}+ \Gt_{\omega} q, \, \fhi^{\la} \in{\mathscr D} ( H_0^{\omega} ), q\in {\mathscr D} (\galao ), \lf( \al  +  \galao \:  \ri)q = \I \,\fhi \ri\},
	\end{equation}
	\begin{equation}
	\label{hamiltonian3d}
		( H_\al^{\omega} +\la ) u = ( H_0^{\omega} +\la )\fhi^{\la},
	\end{equation}
	and the resolvent of $H_{\al}$ can be represented as
	\begin{equation}
	\label{resolvent3d}
		 \lf( H_{\al}^{\omega} + \la \ri)^{-1} f = G^{\la}_{\omega} f  + \Gt_{\omega}  q_f ,
	\end{equation}
	where, for any $ f \in L^2(\erre^6) $, $q_f$ is a solution of
	\begin{equation}
	\label{eqresolvent3d}
		\lf( \al  +  \gala_{\omega} \:\ri) q_f  = \I \, G^{\la}_{\omega} f .
	\end{equation}
\end{teo}

\nt
The operators \eqref{hamiltonian3d} give rise to a one-parameter family of self-adjoint operators, which actually coincides with a family of self-adjoint extension of the three-dimensional analogous of the operator $ \widetilde{H_0} $ introduced in the previous Section. Note that the free Hamiltonian $ H_0^{\omega} $ belongs to the family and it is given by \eqref{hamiltonian3d} for $ \al = + \infty $, exactly as in the two-dimensional case.

\subsection{Spectral Analysis}

\nt
Most of the results proved in the two-dimensional case apply also to the three-dimensional one and there are only minor differences in the proofs. Hence we shall often omit the details and refer to the two-dimensional case. 
\newline
The spectral properties of $ H_{\al}^{\omega} $ are strictly related to spectral properties of the operator $ \gala_{\omega} $, so we shall start by studying the latter: 
\spazietto

\begin{prop}[Spectral Analysis of $ \galao $]
	\label{gala spectrum 3d}
	\mbox{}	\\
	The domain $ \DFia $ can be characterized in the following way:
	\beq
	\label{Gamma domain 3d} 
		\DFia = \left\{ q\in L^2(\erre^3) \: \Big| \: q \in \mathcal{H}^{1/2}(\erre^3)\, , \,
		\hat{q} \in \mathcal{H}^{1/2}(\erre^3) \right\}.
	\eeq   	
	On this  domain $\Fia$ is closed and defines a self-adjoint operator  $ \galao $. For any $ \la >0$ the spectrum $ \sigma(\galao) $ is purely discrete, i.e., $ \sigma(\galao) = \sigma_{\mathrm{pp}}(\galao) $. 
	\newline
	Let $ \ga_n(\la) $, $ n \in \enne $, be the eigenvalues of $ \gala_1 $ arranged in an increasing order 
	$( \lim_{n \rightarrow \infty} \ga_n(\lambda) = + \infty )$. 
	For every $ n \in \enne $, $ \ga_n(\la) $ is a non-decreasing function of $ \la $. Furthermore $ \lim_{\la \rightarrow 0} \ga_0(\la) = -\infty $ and the other eigenvalues remain bounded below, i.e.,  for any $ \la \geqslant 0 $, there exists a finite constant $ c $  such that $ \ga_n(\la) \geqslant - c $, for any $ n \in \enne$, $ n > 0 $. 
\end{prop} 

\begin{dem}	
	Let us set again $ \omega = 1 $ and denote by $ \gala $ the operator $ \gala_1 $.
	\newline
	We can decompose $ \gala = a^{\la} + \gala_0 $, where $ \gala_0 $ is the self-adjoint operator associated with the positive quadratic form
	\beq	
		\Phi^{\la}_0[q] \equiv \half \int_{\erre^2} d\xv d\xvp \:  G^{\la}(\xv,\xv ; \xvp,\xvp) \absq{ q(\xv) - q(\xvp) }.
	\eeq

\n
	Since $ \Phi_0 $ is positive and $ a^{\la}(x) $ is an unbounded function, which is however bounded below for any $ \la > 0 $,  
	$ \gala $ is an unbounded operator which is bounded below. 
	Notice that $ a^{\la}(x) $ is a monotone increasing function of $x$ and $a^{\la}(x)  \simeq c x$ 
	for $x \ten \infty$; furthermore $ a^{\la}(x)  \geqslant  a^{\la}(0) $, $a^{\la}(0)$ is a monotone 
	increasing function of $\la$ and $a^{\la}(0) \simeq c\sqrt{\la} $ for $\la \ten \infty$. Hence 
	by the lower bound \eqref{inferiore} we have that 
	for any
	$\al \in \erre$, there exists $\la_0 > 0 $ such that for $\la > \la_0$ the quadratic form $ \Phi_{\al}^{\la}$ is positive and 
	bounded below; for such $\la$ the operator $\gala$ is invertible.
	
	\n
	The claim on the spectrum of $\gala$ can be proved in the same way as the two dimensional case, then it is sufficient
	to prove that $ \Phi_{\al}^{\la}$ can be written in the following way:
	\beq
		\Phi_{\al}^{\la}[q] = \int_{\erre^2} d\vec{k} \: \left( \al + \tilde{a}^{\la}(k) \right) \absq{\hat{q}(\vec{k})} + \half \int_{\erre^4} d\vec{k} d\vec{k}^{\prime} \:  \tilde{G}^{\la}(\vec{k} ; \vec{k}^{\prime}) \absq{ \hat{q}(\vec{k}) - \hat{q}(\vec{k}^{\prime}) },
	\eeq 
	where 
	\beq
		\tilde{a}^{\la}(k) \equiv \frac{1}{4\pi} \lf\{ C + \int_0^1 d\nu \: \frac{1}{1-\nu} \lf[1 - \frac{4 \nu^{\la-1} (1-\nu)}{(1+\nu^2)\ln \frac{1}{\nu} + 1 - \nu^2} \exp \lf( - \frac{(1-\nu^2)\ln \frac{1}{\nu}}{2 \lf[ (1+\nu^2) \ln \frac{1}{\nu} + 1 - \nu^2 \ri]} k^2 \ri) \ri] \ri\},
	\eeq
	\begin{multline}
			 \tilde{G}^{\la}(\vec{k} ; \vec{k}^{\prime}) \equiv \frac{1}{2\pi^2} \int_0^1 d\nu \: \frac{\nu^{\la-1}}{(1-\nu^2) \ln \frac{1}{\nu} + 2(1-\nu)^2} \cdot	\\
		\cdot \exp \lf\{ - \frac{\lf[ (1+\nu^2) \ln \frac{1}{\nu} + 1 - \nu^2 \ri] \lf( k^2 + {k^{\prime}}^2 \ri)}{2 \lf[(1-\nu^2) \ln \frac{1}{\nu} + 2(1-\nu)^2 \ri]} - \frac{\lf[ 1 - \nu^2 + 2\nu \ln \frac{1}{\nu} \ri]  \vec{k} \cdot \vec{k}^{\prime}}{(1 - \nu^2) \ln \frac{1}{\nu} + 2(1 -\nu)^2} \ri\}.
	\end{multline}
The function $ \tilde{a}^{\la}(k) $ has the same asymptotic behavior for $ k \rightarrow \infty $ as $ a^{\la}(x) $, namely $\hat{a}^{\la}(k) \simeq  c k$ and by applying Rellich's criterion, we have that $\gala$ has pure 
point spectrum.

	\n
	Notice that also the following bound holds:
	\begin{equation}
	\Phi^{\la}_0[q] \leqslant c \| q\|^2_{ {\mathcal H}^{ 1/2 }(\erre^3)  }.
	\label{bakumatsu1}
	\end{equation}
	Indeed, using the inequality in \eqref{tomoe},
	\begin{equation*}
	G^{\la}(\xv,\xv ; \xvp,\xvp) \leqslant c 	
	 \int_0^1 d\nu \: \frac{\nu^{ \la - 1 }}{\lf( \lf( 1- \nu^2 \ri) \ln \frac{1}{\nu} \ri)^{3/2}} 
	 \exp \lf\{  -   \frac{ (\xv-\xvp)^2}{ 2\ln {\frac{1}{\nu}  }  } - \frac{ \nu  ( \xv - \xvp )^2}{ 1 - \nu^2}  \ri\},
\end{equation*}
and taking the Fourier transform, we have
\begin{equation}
\Phi^{\la}_0[q] \leqslant c \int_{\erre} d\vec{k} 
 \int_0^1 d\nu \:	\frac{\nu^{ \la - 1 }}{ \lf( 2\nu \ln \frac{1}{\nu} + 1- \nu^2\ri)^{3/2} } 
\lf\{ 1-  \exp \lf[  - \frac{ (1- \nu^2) \ln \frac{1}{\nu} }{ 2(  1- \nu^2 + 2\nu \ln \frac{1}{\nu} ) } k^2 \ri] \ri\} \: |\hat{q}(k) |^2 .
 \end{equation}
Since
\begin{equation*}
  0\leqslant \int_0^1 d\nu \:	\frac{\nu^{ \la - 1 }}{( 2\nu \ln \frac{1}{\nu} + 1- \nu^2)^{3/2} } 
\lf\{1-  \exp \lf[  - \frac{ 2(1- \nu^2) \ln \frac{1}{\nu} }{ 4( 2\nu \ln \frac{1}{\nu} + 1- \nu^2) } k^2 \ri] \ri\}
 \leqslant c \Mk, 
 \end{equation*}
 \eqref{bakumatsu1} is proved.

\n
Therefore, taking into account the behavior of $a^{\la}$, 
 \eqref{bakumatsu1} implies that there exist $\la_0 > 0$ such that for $\la > \la_0$ we have
 \begin{equation}
 \gala \leqslant c \lf(  \Mx^{1/2} +  \Mp^{1/2} +  \la^{1/2} \ri).
 \label{pigneto1}
 \end{equation}

\n
Also in the three dimensional case the lower bound \eqref{lowerbene} holds as well, but let us
stress that $a^{\la}$ and $\tilde{a}^{\la}$ have different behavior; in particular in the three dimensional case \eqref{Gamma domain 3d} holds true, due to \eqref{pigneto1}.

\n
Monotonicity of the eigenvalues and unboundedness from below of $ \ga_0(\la) $, as $ \la \to 0 $, can be shown exactly as in the two-dimensional case. Note that one has to evaluate the form on the ground state of the three-dimensional harmonic oscillator. 
	\newline
	Furthermore we have the lower bound,
	\bdm
		\bra{q^{\perp}} \gala \ket{q^{\perp}} \geqslant \frac{1}{(4\pi)^{\frac{3}{2}}} \lf[ \frac{1}{2 } + \int_0^{\frac{1}{e}} d\nu \: \frac{1}{(1-\nu)^{\frac{3}{2}}} \ri] \lf\| q^{\perp} \ri\|^2 - \int_{\erre^6} d\vec{x} d\xvp \:  G_{1/e}^{\la}(\vec{x},\vec{x} ; \xvp,\xvp) \: \lf( q^{\perp}(\vec{x}) \ri)^* q^{\perp}(\xvp), 
	\edm
	but, acting as in the proof of \eqref{diffresharm}, we get
	\bdm
		\int_{\erre^6} d\vec{x} d\xvp \:  G_{1/e}^{\la}(\vec{x},\vec{x} ; \xvp,\xvp) \: \lf( q^{\perp}(\vec{x}) \ri)^* q^{\perp}(\xvp) \leqslant \frac{1}{(2\pi)^{\frac{3}{2}}} \lf[ \bra{q^{\perp}} (\Hro+\la-3/2)^{-1} \ket{q^{\perp}} +  2 \lf\| q^{\perp} \ri\|^2 \ri],
	\edm
	so that, if $ q^{\perp} $ is a normalized function orthogonal to the ground state of the harmonic oscillator, $ \bra{q^{\perp}} \gala \ket{q^{\perp}} \geqslant -c $, for some finite constant $ c $.
\end{dem}

\nt
The discrete spectrum of $ H_{\al}^{\omega} $ can now be fully characterized:
\spazietto

\begin{teo}[Negative Spectrum of $H_{\al}^{\omega}$]
	\mbox{}	\\
	For any $ \al \in \erre $, the discrete spectrum $ \sigma_{\mathrm{pp}}(H_{\al}^{\omega}) $ of $ H_{\al}^{\omega} $ is not empty and it contains a number $ N_{\omega}(\al) $ of negative eigenvalues $-E_0(\al,\omega)\leqslant -E_1(\al,\omega) \leqslant \ldots \leqslant 0$ satisfying the scaling
	\beq
	\label{scaling3d}
		E_n(\al,\omega) = \omega E_n(\al/\sqrt{\omega}, 1). 
	\eeq
	The corresponding eigenvectors are given by $ u_n = {\mathcal G}^{E_n}_{\omega}  q_n $, where $ q_n $ is a solution of the homogeneous equation $ \al q_n + \Gamma^{E_n}_{\omega} q_n = 0 $.
	\newline
	Moreover there exists $ \alpha_0 \in \erre $ such that, if $ \al > \al_0 $, $ N_{\omega}(\al) = 1 $ and, for fixed $ \omega $ and $ \al \to -\infty $, $N_{\omega}(\al) \simeq c |\al |^6$. 
	\newline
	The ground state energy has the following asymptotic behavior for fixed $ \omega $: $E_0 \simeq c \al^{-1} $ for $\al \to +\infty$ and $E_0 \simeq c \al^{2} $ for $\al \to -\infty$.
\end{teo}

\begin{dem}
See the proof of Theorem \ref{discrete spectrum 2d}; notice that in the argument used to estimate the asymptotics of $N_{\omega}(\al)$, the spectral distribution of the square root of the 
three-dimensional harmonic oscillator is involved.
\end{dem}

\nt
An interesting consequence of the above Theorem is the existence of a bound state for any $ \al $
and $0< \omega <\infty$, in particular even if $ \al > 0 $ and there are no bound states for the
``reduced" system (we shall come back to this question in the next Section). 
\newline
The asymptotics for $ \omega \to 0 $ and $ \al > 0 $  is exactly as in the one and two-dimensional case, whereas the behavior for $ \omega \to 0 $ and $ \al < 0 $ proves to be much more complicated, due to the ground state asymptotics (see Theorem above). If $ \omega \to \infty $ we expect that the asymptotics depend on a crucial way on the sign of $ \al $, since at least one bound state should survive if $ \al < 0 $, whereas, if $ \al > 0 $, all bound states should disappear in the limit.

\n
Now we shall give a partial characterization of the positive spectrum of
\eqref{ham}, but we first state a result analogous to Lemma \ref{puzzola}:
\spazietto

\begin{lem}
Let us consider $T^k_{\omega} \equiv \Gt_{\omega} ( G^{\la}_{\omega} )^k \Gta_{\omega} $, $k\in \enne $. Then, if $ \la > k $,  $T^k_{\omega} \in 
{\mathscr B}_{p} ( L^2 (\erre^3),L^2(\erre^3))$, for any $p>3 (k+1/2)^{-1} $.
\label{puzzola2}
\end{lem}
\begin{dem}
The proof follows exactly the proof of Lemma \ref{puzzola} and is omitted for the sake of brevity.
\end{dem}

\spazietto
\begin{teo}[Positive Spectrum of $H_{\al}^{\omega}$]
	\mbox{}	\\
	The essential spectrum of $ H_{\al}^{\omega} $  is equal to $ [0,+\infty) $ and the wave operators $\Omega_{\pm}(H_{\al}^{\omega},H_0^{\omega})$ exist and are complete.
\end{teo}

\begin{dem} 
	We shall omit the dependence on $\omega$ for brevity. It is sufficient to prove that $(H_{\al} +\la )^{-1} - (H_{0} +\la )^{-1}$ is a compact operator
	and that $[(H_{\al} +\la )^{-1}]^4 - [(H_{0} +\la )^{-1}]^4$ is trace class
	for some $\la>0$, then the thesis follows from Weyl's Theorem (see Theorem XIII.14 in \cite{rs4} and 
	Corollary 3 of Theorem XI.11 in \cite{rs3}). 

\n
We shall fix $\la$ sufficiently large such that $(\Gamma^{\la} + \al )^{-1}$ exists. Boundedness of  $(\Gamma^{\la} + \al )^{-1}$, H\"{o}lder inequality and the fact that $ \Gta \Gt \in {\mathscr B}_p(L^2(\erre^6),L^2(\erre^6))$, $ p > 6 $, because of Lemma \ref{puzzola2}, imply compactness of $(H_{\al} +\la )^{-1} - (H_{0} +\la )^{-1}$, as in the two-dimensional case. Besides one can show that $ \Gt $ belongs to $ {\mathscr B}_p(L^2(\erre^6),L^2(\erre^3))$, $ p > 12 $, and $ \Gta \in  {\mathscr B}_p(L^2(\erre^3),L^2(\erre^6))$ for the same $ p $. Finally some tedious but straightforward calculations show that
\begin{multline}
[(H_{\al} +\la )^{-1}]^4 - [(H_{0} +\la )^{-1}]^4 =
\Gta (\Gamma^{\la} + \al )^{-1} \Gt \lf( G^{\lambda} \ri)^3 +  
G^{\lambda}\Gta (\Gamma^{\la} + \al )^{-1} \Gt \lf( G^{\lambda} \ri)^2+ \\
\lf( G^{\lambda} \ri)^2 \Gta (\Gamma^{\la} + \al )^{-1} \Gt G^{\lambda}+
\lf( G^{\lambda} \ri)^3\Gta (\Gamma^{\la} + \al )^{-1} \Gt+
\lf(  \Gta (\Gamma^{\la} + \al )^{-1} \Gt       \ri)^2 \lf( G^{\lambda} \ri)^2+ \\
\Gta (\Gamma^{\la} + \al )^{-1} \Gt G^{\lambda} \Gta (\Gamma^{\la} + \al )^{-1} \Gt G^{\lambda} +
\Gta (\Gamma^{\la} + \al )^{-1} \Gt \lf( G^{\lambda} \ri)^2 \Gta (\Gamma^{\la} + \al )^{-1} \Gt + \\
\lf( G^{\lambda} \ri)^2 \lf(  \Gta (\Gamma^{\la} + \al )^{-1} \Gt       \ri)^2 +
G^{\lambda}\Gta (\Gamma^{\la} + \al )^{-1} \Gt  G^{\lambda} \Gta (\Gamma^{\la} + \al )^{-1} \Gt +\\
G^{\lambda}\lf(  \Gta (\Gamma^{\la} + \al )^{-1} \Gt       \ri)^2 G^{\lambda} +
\lf(  \Gta (\Gamma^{\la} + \al )^{-1} \Gt       \ri)^3 G^{\lambda} + 
\lf(  \Gta (\Gamma^{\la} + \al )^{-1} \Gt       \ri)^2 G^{\lambda} \Gta (\Gamma^{\la} + \al )^{-1} \Gt + \\
 \Gta (\Gamma^{\la} + \al )^{-1} \Gt G^{\lambda} \lf(  \Gta (\Gamma^{\la} + \al )^{-1} \Gt       \ri)^2  +
G^{\lambda} \lf(  \Gta (\Gamma^{\la} + \al )^{-1} \Gt       \ri)^3  + 
\lf(  \Gta (\Gamma^{\la} + \al )^{-1} \Gt       \ri)^4,
\label{chepalle}
\end{multline}
and applying Lemma \ref{puzzola2} with $ k = 3 $ to each term, we get the result.
\end{dem}

\section{Conclusions and Perspectives}

\nt
We have studied a quantum system composed of a test particle and a harmonic oscillator interacting through a zero-range force. We have given a rigorous meaning to the hamiltonian $ H_{\al}^{\omega} $ of the system, described the properties of its spectrum and established asymptotic completeness for the scattering operators $ \Omega_{\pm}(H_{\al}^{\omega},H_0^{\omega}) $, where $ H_0^{\omega} $ is the hamiltonian of the system without the zero-range force.
\nw
We remark that our analysis could be useful in an explicit  multichannel scattering approach, the channels being labeled by the bound states of the harmonic oscillator.
\nw
The negative part of the spectrum of $ H_{\al}^{\omega} $ for $\omega>0$ is discrete and we have given estimates of the number of bound states. There is a peculiar feature of this part of the spectrum: in the three dimensional setting, in the case of a fixed center, i.e. $\omega=\infty$, when the parameter $ \alpha $ is negative, there is exactly one bound state, while, in the case $ \alpha > 0 $, the spectrum is absolutely continuous. In our case, if $ \alpha > 0 $, there is always a bound state and, if $ \alpha < 0 $, the number of bound states increases as the strength of the harmonic force goes to zero.
\nw
We might interpret this feature as  due to the fact that bound states of the harmonic oscillator provide a mechanism through which the test particle is bound, even if the interaction due to the zero-range force is ``repulsive''.
\nw
In spite of the simplicity of the system, the analysis shows the presence of at least some of the difficulties of the more general case in which the test particle interacts with $N$ harmonic oscillators (Rayleigh gas); we shall discuss this more general case in a forthcoming paper.

\spaz
\nt
\textbf{Acknowledgments:} The authors are very grateful to A. Teta for many interesting and helpful discussions.


\begin{thebibliography}{99}

\bibitem[AGH-KH]{al} \textsc{S. Albeverio, F. Gesztesy, R. Hogh-Krohn, H. Holden}, {\em Solvable
Models in Quantum Mechanics}, Springer-Verlag, New-York, 1988.
\bibitem[BC]{bc} \textsc{J. Bellandi, E.S. Caetano Neto}, The Mehler Formula and the Green Function of Multidimensional Isotropic Harmonic Oscillator, {\em J. Phys. A: Math Gen.} {\bf 9} (1976), 683-685.
\bibitem[BF]{bf} \textsc{F.A. Berezin, L.D. Faddeev}, A Remark on Schrodinger Equation with a Singular
Potential, {\em Sov. Math. Dokl.} {\bf 2} (1961), 372-375.
\bibitem[D]{da} \textsc{A.S. Davydov}, {\em Quantum Mechanics}, Pergamon Press, Oxford, 1965.
\bibitem[DFT]{me} \textsc{G. Dell'Antonio, D. Finco, A. Teta}, Singularly Perturbed Hamiltonians of a Quantum Reyleigh Gas Defined as Quadratic Forms, {\em Pot. Anal.} {\bf 22} (2005), 229-261.
\bibitem[DO]{d} \textsc{Y.N. Demkov, V.N. Ostrovsky}, {\em Zero-range Potentials and their Applications in Atomic Physics},
Plenum Press, New York and London, 1988.
\bibitem[F]{fe} \textsc{E. Fermi}, Sul Moto dei Neutroni nelle Sostanze Idrogenate, (in italian) {\em Ricerca Scientifica} {\bf 7} (1936), 13-52.
\bibitem[HS]{hs} \textsc{P.R. Halmos, V.S. Sunder}, {\em Bounded Integral Operators on $L^2$ Spaces}, Springer Verlag, New York, 1978.
\bibitem[K]{k} \textsc{T. Kato}, {\em Perturbation Theory for Linear Operators}, Classics in Mathematics, Springer-Verlag, Berlin, 1995.
\bibitem[KP]{kp} \textsc{R. Kronig, W.G. Penney}, Quantum Mechanics of Electrons in Crystal Lattices, {\em Proc. R. Soc. A} {\bf 130} (1931), 499-513.
\bibitem[LA]{la} \textsc{L.D. Landau, E.M. Lifshitz}, {\em Non Relativistic Quantum Mechanics}, Mir, Moscow, 1967.
\bibitem[LO]{lo} \textsc{S.W. Lovesey}, {\em Theory of Neutron Scattering from Condensed Matter}, Clarendon Press, Oxford, 1984.
\bibitem[S]{si} \textsc{B. Simon}, {\em Trace Ideals and Their Applications}, Mathematical Surveys and Monographs \textbf{120}, American Mathematical Society, Providence, RI, 2005.
\bibitem[RSI]{rs1} \textsc{M. Reed, B. Simon}, {\em Methods of Modern Mathematical Physics. Vol I: Functional Analysis}, Academic Press, San Diego, 1972.
\bibitem[RSII]{rs2} \textsc{M. Reed, B. Simon}, {\em Methods of Modern Mathematical Physics. Vol II: Fourier Analysis and Self Adjointness}, San Diego, 1975.
\bibitem[RSIII]{rs3} \textsc{M. Reed, B. Simon}, {\em Methods of Modern Mathematical Physics. Vol III: Scattering Theory}, Academic Press, San Diego, 1975.
\bibitem[RSIV]{rs4} \textsc{M. Reed, B. Simon}, {\em Methods of Modern Mathematical Physics. Vol IV: Analysis of Operators}, Academic Press, San Diego, 1978.
\bibitem[Y]{y} \textsc{K. Yosida}, {\em Functional Analysis}, Classics in Mathematics, Springer-Verlag, Berlin, 1995.
\end{thebibliography}
\end{document}